\pdfoutput=1

\documentclass[final,5p,times,twocolumn]{elsarticle}
\usepackage{amssymb}
\newcounter{bla}

\usepackage{acronym}
\usepackage{algpseudocode}
\usepackage{algorithm}
\usepackage{xspace}
\usepackage{amsmath}
\usepackage{graphicx}
\usepackage{xcolor}
\usepackage{booktabs}
\usepackage{array}
\usepackage[leftcaption]{sidecap}
\usepackage{multirow}

\usepackage[binary-units=true,detect-none]{siunitx}
\usepackage[%
  hyperfootnotes=false,%
  pdfpagelabels]{hyperref}
\hypersetup{%
  colorlinks=true, linktocpage=true,%
  breaklinks=true, pdfpagemode=UseNone, pageanchor=true, pdfpagemode=UseOutlines,%
  plainpages=false, bookmarksnumbered, bookmarksopen=true, bookmarksopenlevel=1,%
  hypertexnames=true, pdfhighlight=/O,%nesting=true,%frenchlinks,%
}

\usepackage[capitalize,noabbrev]{cleveref}

\newcommand\mir{MiRheo\xspace}
\newcommand\udx{uDeviceX\xspace}
\newcommand\nv{Nvidia\xspace}

\newcommand\pvs{\textit{Particle Vectors}\xspace}

\newacro{DPD}{Dissipative Particle Dynamics}
\newcommand{\DPD}{\ac{DPD}\xspace}

\newacro{MD}{Molecular Dynamics}
\newcommand{\MD}{\ac{MD}\xspace}

\newacro{RBC}{Red Blood Cell}
\acrodefplural{RBCs}{Red Blood Cells}
\newcommand{\RBC}{\ac{RBC}\xspace}
\newcommand{\RBCs}{\acp{RBC}\xspace}

\newacro{CTC}{Circulating Tumor Cell}
\acrodefplural{CTCs}{Circulating Tumor Cells}

\newcommand{\CTCs}{\acp{CTC}\xspace}

\newacro{SDF}{Signed Distance Function}
\newcommand{\SDF}{\ac{SDF}\xspace}

\newacro{DLD}{Deterministic Lateral Displacement}
\newcommand{\DLD}{\ac{DLD}\xspace}

\newacro{GPU}{Graphics Processing Unit}
\acrodefplural{GPUs}{Graphics Processing Units}
\newcommand{\GPU}{\ac{GPU}\xspace}
\newcommand{\GPUs}{\acp{GPU}\xspace}

\renewcommand{\vec}[1]{\mathbf{#1}}

\newcommand{\Uip}[0]{U_{\text{in-plane}}}
\newcommand{\Ube}[0]{U_{\text{bending}}}
\newcommand{\Uar}[0]{U_{\text{area}}}
\newcommand{\Uvo}[0]{U_{\text{volume}}}

\newcommand{\Atot}{A^{\text{tot}}}
\newcommand{\Aotot}{A^{\text{tot}}_0}
\newcommand{\Votot}{V^{\text{tot}}_0}
\newcommand{\Rey}{\mathrm{Re}}

\graphicspath{{figures/figures/}{./figures/}}
 
\journal{Computer Physics Communications}

\begin{document}

\begin{frontmatter}

%\title{A \LaTeX{} template for CPC Computer Physics Descriptions}
\title{\mir: High-Performance Mesoscale Simulations for Microfluidics}

%\author[a]{First Author\corref{author}}
%\author[a,b]{Second Author}
%\author[b]{Third Author}
%
%\cortext[author] {Corresponding author.\\\textit{E-mail address:} firstAuthor@somewhere.edu}
%\address[a]{First Address}
%\address[b]{Second Address}

\author[a]{Dmitry Alexeev}
\author[a]{Lucas Amoudruz}
\author[a]{Sergey Litvinov}
\author[a]{Petros Koumoutsakos\corref{author}}
\cortext[author] {Corresponding author.\\\textit{E-mail address:} petros@ethz.ch}
\address[a]{Computational Science and Engineering Laboratory, Clausiusstrasse 33, ETH Z\"{u}rich, CH-8092, Switzerland}

\begin{abstract}
%A submitted program is expected to be of benefit to other physicists or physical chemists, or be an exemplar of good programming practice, or illustrate new or novel programming techniques which are of importance to some branch of computational physics or physical chemistry.
%
%Acceptable program descriptions can take different forms. The following Long Write-Up structure is a suggested structure but it is not obligatory. Actual structure will depend on the length of the program, the extent to which the algorithms or software have already been described in literature, and the detail provided in the user manual.
%
%Your manuscript and figure sources should be submitted through the Elsevier Editorial System (EES) by using the online submission tool at \\
%http://www.ees.elsevier.com/cpc.
%
%In addition to the manuscript you must supply: the program source code; job control scripts, where applicable; a README file giving the names and a brief description of all the files that make up the package and clear instructions on the installation and execution of the program; sample input and output data for at least one comprehensive test run; and, where appropriate, a user manual. These should be sent, via email as a compressed archive file, to the CPC Program Librarian at cpc@qub.ac.uk.

The transport and manipulation of particles and cells in microfluidic devices has become a core methodology in domains ranging from molecular biology to manufacturing and drug design. The rational design and operation of such devices can benefit from simulations that resolve flow-structure interactions at sub-micron resolution. We present a computational tool for large scale, efficient and high throughput mesoscale simulations of fluids and deformable objects at complex microscale geometries.
The code employs Dissipative Particle Dynamics for the description of the flow coupled with visco-elastic membrane model for red blood cells and can also handle  rigid bodies and complex geometries.
The  software (\mir) is deployed on hybrid GPU/CPU architectures  exhibiting unprecedented time-to-solution performance and excellent weak and strong scaling for a number of benchmark problems. \mir exploits the capabilities of GPU clusters, leading to speedup of up to 10X in terms of time to solution as compared to state-of-the-art software packages and reaches 90\% -- 99\% weak scaling efficiency on 512 nodes of the Piz Daint supercomputer.
The software  \mir, relies on a Python interface to facilitate the solution of complex problems and it is open source. We believe that \mir constitutes a potent computational tool that can greatly assist studies of microfluidics.

\end{abstract}


\end{frontmatter}

% \linenumbers

% Computer program descriptions should contain the following
% PROGRAM SUMMARY.

\noindent {\bf PROGRAM SUMMARY}
  %Delete as appropriate.

\begin{small}
\noindent
{\em Program Title:} \mir                                         \\
{\em Licensing provisions: GPLv3 }                                   \\
{\em Programming language:} C++, CUDA, Python                                  \\
%{\em Supplementary material:}                                 \\
  % Fill in if necessary, otherwise leave out.
% Seems not applicable, we're not in the CPC library
%{\em Journal reference of previous version: \cite{rossinelli2015b}}                  \\
%{\em Does the new version supersede the previous version? yes}   \\
%{\em Reasons for the new version:} The new version of the code adds support for rigid bodies and multi-phase solvent, as well as improves user experience with simple Python interface\\
%{\em Summary of revisions:} ???\\
{\em Nature of problem:} 3D simulations of microfluidic flows in complex geometries with suspended rigid bodies and deformable membranes such as cells, bacteria and microparticles. \\
  %Describe the nature of the problem here (approx. 50-250 words). \\
{\em Solution method:} Dissipative Particle Dynamics are used to represent the fluid.
Cell membrane dynamics are described through potentials for shear and bending energies  that are discretized on a triangular mesh  and by additional constraints on cell volume and membrane area. The model incorporates membrane viscosity and interactions between membranes and the surrounding fluid. 
Rigid objects and boundaries are represented by groups of particles with prescribed  center of mass and rotation quaternion.
Time integration is performed using the  Velocity-Verlet algorithm.  \\
  %Describe the method solution here. (approx. 50-250 words)
{\em Additional comments including Restrictions and Unusual features:} The code runs on \nv GPU accelerators starting with the Kepler generation.\\
  %Provide any additional comments here. (approx. 50-250 words)
   \\
%
%\begin{thebibliography}{0}
%\bibitem{1}Reference 1         % This list should only contain those items referenced in the                 
%\bibitem{2}Reference 2         % Program Summary section.   
%\bibitem{3}Reference 3         % Type references in text as [1], [2], etc.
%                               % This list is different from the bibliography at the end of 
%                               % the Long Write-Up.
%\end{thebibliography}

%* Items marked with an asterisk are only required for new versions
%of programs previously published in the CPC Program Library.\\
\end{small}

\section{Introduction}
%{\color{red}
%\begin{itemize}
%	\item Blood flow and microfluidics are cool and important.
%	\item Simulations are good and important.
%	\item DPD simulation are good and important. Examples of recent DPD blood things: Pivkin, Fedosov, Karniadakis
%	\item Most of works don't disclose code and/or pretty slow. Solution - go for fast gpu stuff.
%	\item A few words about gpus and why they are cool
%	\item here we present mirheo -- a successor of udx
%	\item list of features
%\end{itemize}
%}

Microfluidic devices are used to transport, control, analyse and manipulate nanoliter quantities of liquids, gasses and  other substances ~\cite{Beebe2002,Whitesides2006,Streets2013} in  natural sciences and engineering as well as in clinical research~\cite{Squires2005,Streets2013,Sackmann2014,Zhang2016}. Fluid flows in  microfluidic devices are characterized by Reynolds numbers that are low  ($10^{-3}$ - $10^1$) or moderate ($10^{-1}$ - $10^2$) in high throughput settings. Fluid flows at the microscale have been often modeled by the continuum Navier Stokes equations with various discretizations \cite{Clague2001,DiCarlo2009,Cimrak2012,Karabacak2014,Scherr2015,grimmer2019}.
Continuum models  have been successful in capturing several key features of microscale flows but at the same time face limitations as they are not able to resolve phenomena affected by fluctuations and related biophysical processes~\cite{Freund2014}. Moreover the complex geometries, the multiple deforming objects as well as the need to handle chemistry and related processes often adds computational complexity to the classical Navier Stokes solvers leading to expensive computations with low  time-to-solution~\cite{rahimian2010petascale}.
We note for example that state-of-the-art simulations of \RBCs that use scalable boundary integral methods have only used a few hundred \RBCs in two dimensions~\cite{Quaife2014,Kabacaoglu2018}.

An alternative simulation approach for micorfluidics uses mesoscale \DPD method, which represents the fluids and suspended objects as collections of particles.
\DPD is a stochastic, short-range particle method that bridges the gap between Molecular Dynamics and Navier-Stokes equations~\cite{Espanol1995}.
It has been used extensively to model complex fluids such as colloidal suspensions, emulsions and polymers~\cite{Boek1997,Moeendarbary2010} and has recently become a key method for the study of the blood rheology~\cite{Quinn2011,Zhang2014,Fedosov2014,Lanotte2016}.
However, currently most of these simulations are carried out by non open source software based on LAMMPS~\cite{Plimpton1995} \MD package. LAMMPS is notable for its flexibility and capability to model multiphysics but at the same time this may come at the expense of speed in domain specific settings  such as \MD \footnote{ \href{http://www.hecbiosim.ac.uk/benchmarks}{http://www.hecbiosim.ac.uk/benchmarks} }.
A  number of open source codes\cite{Tang2014,Blumers2017,seaton2013dl_meso},provide  little usage instructions, but at the same time demonstrate  better performance than LAMMPS by exploiting extensively \GPUs to accelerate the most computationally-expensive kernels.
As many problems in computational science are usually data-parallel, the appeal of moving some computations to the \GPU is high~\cite{Schulte2015}: while demanding highly parallel problems and careful implementations, the \GPUs offer unmatched FLOP performance and memory bandwidth, outperforming state-of-the-art server-grade CPUs by a factor of $5$ to $10$.

However, transferring the existing code onto a \GPU is usually not a trivial task due to significant architectural differences in the hardware such as cache size, width of the vector instructions, different control flow penalties etc..
%Often a different data layout is beneficial: while Structure of Arrays (SoA) is almost always beneficial for the CPU due to SIMD instruction, the G
%\GPU hardware adopts Single Instruction -- Multiple Threads (SIMT) paradigm memory hierarchy that may benefit from spatial locality of Array of Structures (AoS).
%
Another aspect of porting the application to the \GPU is the memory traffic through the PCI-E bus between the accelerator and the CPU.
The bus only delivers a fraction of main RAM bandwidth, and if used extensively, may easily result in a performance bottleneck.
Therefore often porting parts of the application onto the \GPU may not be either beneficial nor easy, and starting from scratch is to be preferred.

\begin{figure}[ht]
	\caption { Simulation of red blood cells and microscale drug carriers inside capillaries. }
	\centering
	\includegraphics[width=0.99\columnwidth]{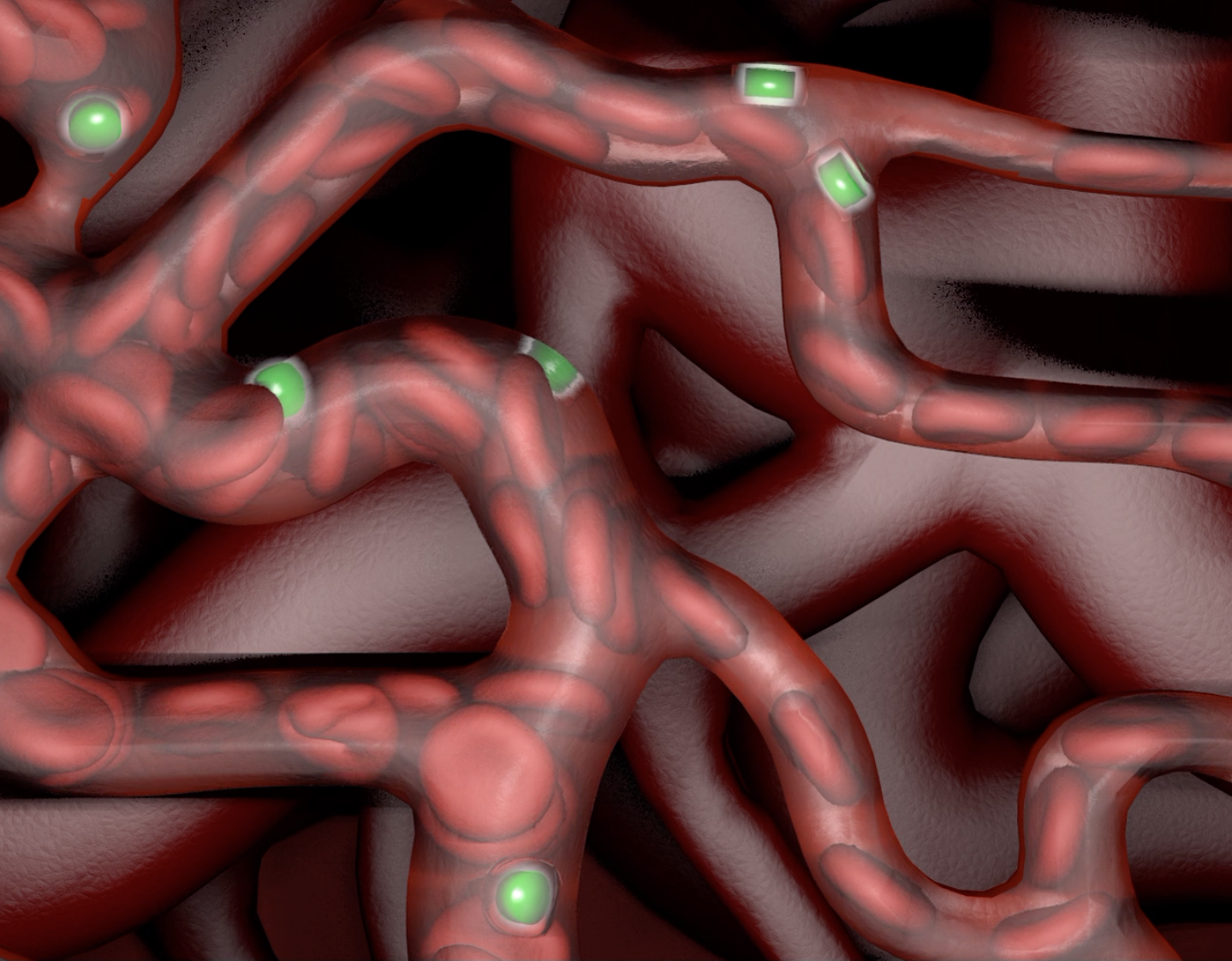}
	\label{fig:capillaries}
\end{figure}

Here we present \mir\footnote{\href{https://github.com/cselab/Mirheo}{https://github.com/cselab/Mirheo} }, a high-throughput software for microfluidic flow simulations in complex geometries with suspended visco-elastic cell membranes and rigid objects, written exclusively for \GPUs and clusters of \GPUs (e.g. see \cref{fig:capillaries}).
It is a successor of \udx code~\cite{Rossinelli2015} with improved performance, usability, extensibility and many additional features.
\mir handles complex geometries, large number of suspended rigid bodies and cells, fluids with different viscosity and provides a flexible yet efficient and well-documented way to specify the simulation setup and parameters.
In the rest of the paper we first introduce the employed numerical method (\cref{sec:numerics}), then go over details of our implementation and parallelization strategies (\cref{sec:implementation}), followed by code validation (\cref{sec:validation}) and benchmarks (\cref{sec:benchmarks}) before concluding (\cref{sec:conclusion}).
% Next we show a simple example of code usage (\cref{sec:manual}), and finally conclude (\cref{sec:conclusion}).
%\mir is fast, outperforming similar state-of-the-art particle codes by roughly $1.5 - 10$ times, scalable, reaching $90 - 99\%$ efficiency on hundreds of nodes, and yet user-friendly, exposing its features via Python interface.

 \section{Numerical method}
\label{sec:numerics}

\mir is based on the \DPD method, which yields fluctuating hydrodynamic~\cite{Espanol1995,Groot1997}.
The software accommodates flows in complex geometrical domains as well as deformable and rigid objects suspended in the fluid. More specifically, the supported objects are visco-elastic closed shells (representing cell membranes discretized on triangular meshes) and rigid bodies of arbitrary shape.
The evolution of the system is governed by pairwise particle forces while enforcing of the no-slip and no-through boundary conditions where applicable.

\subsection{Dissipative particle dynamics}

The \DPD, is a  particle based method introduced by Hoogerbrugge \cite{Hoogerbrugge1992} and further formulated and developed  in ~\cite{Groot1997,Espanol1995}.
In \DPD the fluid is described by a set of particles in the 3D space.
Each particle is characterized by its mass $m$, position $\vec{r}$ and velocity $\vec{v}$.
Particles evolve in time according to the Newton's law of motion:
\begin{equation}
\begin{aligned}
  \frac{d \vec{r}}{dt} &= \vec{v}, \\
  \frac{d \vec{v}}{dt} &= \frac {1}{m} \vec{F},
\end{aligned}
\end{equation}
where $\vec{F}$ is the force exerted on the particle and $t$ is time.
The force fields are usually expressed in terms of the distance $r$ between particles and they imply local interactions as they vanish after a cutoff radius $r_c$.
The particles interact through central forces, which implies, by the Newton's third law, conservation of linear and angular momentum.
The \DPD forces acting on the particle indexed by $i$ are written as
\begin{equation}
\vec{F}_i = \sum\limits_{j} \left( \vec{F}_{ij}^C + \vec{F}_{ij}^D + \vec{F}_{ij}^R \right),
\end{equation}
where the force is composed of a conservative, dissipative and random term.
The conservative term acts as purely repulsive force and reads
\begin{equation}
\vec{F}_{ij}^C = \alpha w(r_{ij}) \vec{e}_{ij},
\end{equation}
where $r_{ij} = |\vec{r}_{ij}|$, $\vec{r}_{ij} = \vec{r}_i - \vec{r}_j$, $\vec{e}_{ij} = \vec{r}_{ij} / r_{ij}$ and
\begin{equation}
w(r) = \begin{cases}
  1 - r/r_c, & \text{if } r < r_c, \\
  0,         & \text{otherwise}.
\end{cases}
\end{equation}
The dissipative and random terms are given by
\begin{equation}
\begin{aligned}
  \vec{F}_{ij}^D &= - \gamma \left( \vec{v}_{ij} \cdot \vec{e}_{ij} \right) w_D(r_{ij}) \vec{e}_{ij},\\
  \vec{F}_{ij}^R &= \sigma \xi_{ij} w_R(r_{ij}) \vec{e}_{ij}.
\end{aligned}
\end{equation}
The random variable $\xi_{ij}$ is independent Gaussian noise satisfying
$\langle \xi_{ij}(t) \xi_{lm}(t') \rangle = \delta(t-t') \left( \delta_{il} \delta_{jm} + \delta_{im} \delta_{jl} \right)$,
$\xi_{ij} = \xi_{ji}$ and $\langle \xi_{ij} \rangle = 0$.  
The parameters $\gamma$ and $\sigma$ are linked through the fluctuation-dissipation relation $w_D = w_R^2$ and $\sigma^2 = 2\gamma k_BT$~\cite{Espanol1995}.
The dissipative kernel has the form $w_R(r) = w^k(r)$ with $s \in (0, 1)$~\cite{Fan2006}.

\subsection{Objects representation} \label{sec:model:objects}

Rigid objects are modeled as groups of particles moving with the same velocity field of their center of mass. Their surface geometry is expressed either analytically or by a triangular mesh-based representation.
%We use quaternions to describe their orientation.
The state of a rigid object is fully determined by its center of mass, orientation (stored as a quaternion), linear and angular velocities.

The visco-elastic incompressible membrane is modeled by a triangular mesh with particles as its vertices.
The elastic potential energy of the membrane with constant volume and area is given by~\cite{Fedosov2010b}:
\begin{equation}
U = \Uip + \Ube + \Uar + \Uvo.
\end{equation}
$\Uip$ accounts for the energy of the elastic spectrin network of the membrane, including an attractive worm-like chain potential and a repulsive potential such that a nonzero equilibrium spring length can be obtained.
\begin{equation}
\Uip = \sum\limits_{j = 1}^{N_e}
\left[
  \frac {k_s l_m \left( 3x_j^2 - 2x_j^3 \right)}{4(1-x_j)}
  +
  \frac{k_p}{l_0}
  \right],
\end{equation}
where $k_s$ is the spring constant, $x_j$ is the normalized spring length and $N_e$ is the number of mesh edges.
The bending energy term, $\Ube$, models the resistance of the lipid bilayer to bending.
We implement two different energy models for the membrane dynamics. The first is attributed to Kantor and Nelson~\cite{kantor1987phase}:
\begin{equation}
\Ube^{KN} = \sum\limits_{j = 1}^{N_s} k_b \left[  1-\cos(\theta_j - \theta_0) \right],
\end{equation}
where $k_b$ is the bending constant, $\theta_j$ is the angle between two adjacent triangles (called dihedral) and $\theta_0$ is the equilibrium angle.
The second was developed by  J\"ulicher~\cite{julicher1996morphology}:
\begin{equation}
  \Ube^{J} = 2 k_b \sum\limits_{j = 1}^{N_v} \frac{M_j^2}{A_j},
\end{equation}
where
\begin{equation*}
  M_j = \frac{1}{4} \sum\limits_{\langle k,n \rangle}^{(j)} l_{kn} \theta_{kn}.
\end{equation*}
$\Uar$ and $\Uvo$ are penalization terms accounting for area and volume conservation of the membrane:
\begin{equation}
\begin{aligned}
  \Uar &= \frac{k_a (\Atot-\Aotot)^2}{2\Aotot} + \sum\limits_{j = 1}^{N_t} \frac{k_d (A_j-A_0)^2}{2A_0}, \\
  \Uvo &= \frac{k_v(V-\Votot)^2}{2\Votot},
\end{aligned}
\end{equation}
where $A_j$ is the area of a single triangle, $\Atot = \sum_{j = 1}^{N_t} {A_j}$, $V$ is the volume enclosed by the membrane and $N_t$ is the number of triangles in the mesh.

The membrane viscosity is modeled by an additional pairwise interaction between particles sharing the same edge.
This interaction contains a dissipative and random term with the same form as the \DPD interaction with $w^R(r) = 1$.

\subsection{Boundary conditions}

Solid boundaries in the computational domain are represented via a \SDF, zero value isosurface which defines the wall surface.
A layer of frozen particles with thickness of $r_c$ is located just inside the boundary.
The no-through condition on the wall surface is enforced via a bounce-back mechanism~\cite{Revenga1998}.
These particles have the same radial distribution function as the fluid particles, and interact with the latter with the same \DPD forces.
This ensures the no-slip condition as well as negligible density variations of the fluid in proximity to the wall~\cite{fedosov2008BC, Kotsalis2009, Werder2005}.

The fluid-structure interactions for the rigid objects are similar to the ones employed for the walls.
The surface impenetrability is ensured by bouncing-back solvent particles off the rigid objects surfaces with linear and angular momentum conservation.
%An extra force and torque due to bounce-back is applied to the bodies to ensure momentum conservation. 

In order to maintain the no-slip and no-through flow boundary conditions on the membrane surface, we use the technique originally proposed in ~\cite{Fedosov2010b}.
We assume that a membrane is always surrounded by fluid from both sides, with the same density and conservative potential, but the code allows for different fluid viscosities.
We then let the fluid particles across the membrane interact only with the conservative part of the \DPD force, and in contrast, make the fluid--membrane interaction purely viscous. 
In that way we maintain constant radial distribution function in the liquids in proximity of the membrane, and with the appropriate choice of the liquid--membrane viscous parameter the no-slip condition is satisfied.
The no-through condition is also enforced via the bounce-back mechanism.

\section{Implementation details}
\label{sec:implementation}

The outline of \mir shares several key features with  classical \MD application with local interactions. However the introduction of the bounce-back mechanism, adoption of relatively large time-step and the necessity to operate on membranes consisting of hundreds and thousands of particles requires extra attention that distinguishes \mir from classical \MD implementations.
The software targets \GPU-enabled clusters and use established technologies and libraries such as C++, CUDA, MPI, HDF5 and Python.

\begin{figure}[ht]
	\caption { Layout of the main \mir components within one node of Piz Daint supercomputer. Interconnect (Infiniband) is in red, CPU part shaded in blue, \GPU part -- in green. Only the postprocess task performs heavy I/O. }
	\centering
	\includegraphics[width=0.99\columnwidth]{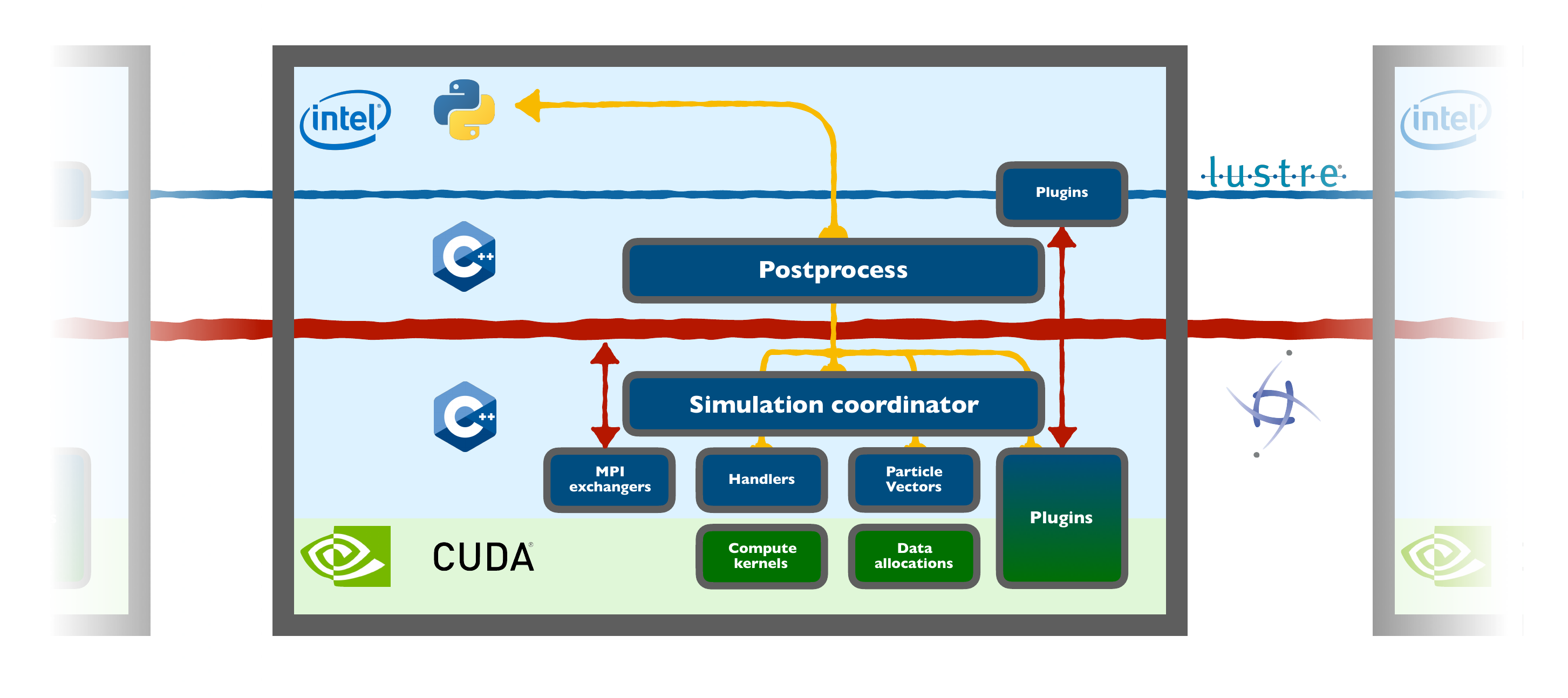}
	\label{fig:scheme}
\end{figure}

\subsection{Algorithmic overview}

The majority of design decisions in \mir have been dictated by the demands of the \GPU architectures and the requirement for a robust and extensible code.
The overall structure of \mir includes the following main components (see \cref{fig:scheme}):
\begin{itemize}
	\item Data management classes, called \pvs. They store particles of a specific type and their properties.
	Objects, like rigid bodies or cell membranes, are also implemented as \pvs for uniformity.
	\item Various handler classes that implements various actions on the \pvs, e.g., integration, force computations, wall interactions.
	\item Plugins, which provide a convenient and non intrusive way of adding functionalities to \mir.
	\item Coordinator classes, that perform initial simulation setup and time-stepping.
	These classes stitch together the \pvs, handlers and plugins into an extensive simulation pipeline.
	\item Python bindings, that provide a way to create and manipulate the data and handlers.
\end{itemize}

We use MPI parallelization by employing domain decomposition into equal rectangular boxes, such that each MPI rank keeps only the local particles. 
To reduce communication between the MPI ranks, we assign all the particles of a single object to a single MPI process depending on the center of mass of the object.
The core data (including particles, cell lists, forces, etc.) are stored in the \GPU RAM, while only the objects and particle adjacent to the subdomain boundaries require to be transferred to the CPU memory and communicated via MPI to the adjacent ranks.
Moreover, we organize the particle data as Structure-Of-Arrays (SOA) in order to optimize memory traffic, and also to allow dynamic addition of extra properties per each particle or each object.

The time-stepping pipeline of \mir is organized as follows:
\begin{enumerate}
\item Create the cell-lists for all the types of particles (\pvs) and interaction cut-offs.
	In a typical simulation the cut-off radius is the same for all the pairwise forces involved and we found that using a separate cell-list for different cut-off radii is beneficial in terms of overall performance.
\item Using the created cell-lists, we identify the \textit{halo} (or \textit{ghost}) particles, that have to be communicated to the adjacent processes.
	The transfer itself is overlapped with the subsequent force computation.
	We will give more details about the overlap in the later section.
\item Compute forces due to the local particles.
\item After the halo exchange is completed, we compute the forces in the system due to particles coming from neighboring processes.
\item Integrate the particles with the fused Velocity-Verlet.
%	\begin{equation}
%		\begin{aligned}
%			\mathbf{v}_p^{n+1/2} & \gets \mathbf{v}_p^{n-1/2} + \frac{\mathbf{f}_n}{m} \delta t \\
%			\mathbf{x}_p^{n+1} & \gets \mathbf{x}_p^n + \mathbf{v}^{n+1/2} \delta t
%		\end{aligned}
%	\end{equation}
	The rigid bodies need special treatment: we first calculate the total force and torque for each body, and then integrate their positions and rotational quaternions.
\item Bounce the particles off the walls, rigid bodies and membranes.
	The forces due to the bounce are saved in the objects and transferred to the next time-step.
\item Identify the particles and objects that have left the local subdomain and send them to the corresponding adjacent MPI rank.
\end{enumerate}

\subsection{Pairwise interactions and cell-lists}

The nominal cost of  computing all the pairwise forces in a system with $N$ particles scales as $\mathcal{O}(N^2)$.
However, in \DPD the pairwise forces only affect the local neighborhood of each particle as the potential vanishes with the increased particle distance and the cost of force calculation is reduced to $\mathcal{O}(N)$.
A common approach to restrict force computation to the particle pairs within a distance $r_{cut}$ is to use the Verlet, or neighbor lists (LAMMPS~\cite{Plimpton1995}, NAMD~\cite{Phillips2005}, GROMACS~\cite{Berendsen1995}, HOOMD-blue~\cite{Anderson2008}).
Such lists store for each particle the indices of all the other particles in the system within the distance $r_{cut} + \varepsilon$.
The non-zero $\varepsilon > 0$ is introduced as the cost of building that structure is significant compared to the force evaluation, therefore it is advisable to rebuild it only once every few time-steps.
For each simulation there exist an optimal $\varepsilon$ that is governed by a balance between building the neighbor list (benefits from large $\varepsilon$) and force evaluation (benefits from smaller $\varepsilon$).

Another distinguishing feature of the \DPD forces with respect to classical \MD force fields such as Lennard-Jones, is the fact that the potential of the conservative force $F^C$ increases at a far lower rate with the pairwise distance (a "soft" potential).
In turn, the typical \DPD time-steps are much larger  than ones used in \MD.
These two factors together result in fast changes of the particle neighborhoods, that in turn would result in frequent Verlet list rebuilding and consequently a performance penalty.

In \mir we use the cell-lists to accelerate the force computation.
We first split the domain of interest into cubic cells with edge length $r_{cut}$, forming a uniform Cartesian grid.
Then the cell-list data structure is defined as a two-way mapping of a particle onto a unique cell.
The particle-cell mapping is trivial and can easily be computed from the particles coordinates.
The construction of the inverse mapping, the cell-list itself, requires the particles to be sorted according to the index of the cell they belong to, and computing the positions in the particle array corresponding to each cell.
%Device atomic operations are heavily used for, but for the typical particle densities of $1-20$ per cell the atomic latencies are effectively hidden.
%
%\begin{algorithm}
%\label{lst:celllist}
%	\begin{algorithmic}
%		\Procedure{CellList}{particles, nx, ny, nz}
%		\State cellStarts $\gets $ Zeros($nx \times ny \times nz + 1$)
%		\State cellSizes  $\gets $ Zeros($nx \times ny \times nz$)
%		\ForAll { $p \in $ particles }
%			\State cellSizes[\Call{ComputeCellId}{$p$}]++
%		\EndFor
%		\State cellStarts $\gets$ \Call{ExclusiveScan}{cellSizes}
%		\State cellSizes  $\gets $ Zeros($nx \times ny \times nz$)
%		\ForAll { $p \in $ particles }
%			\State $cell =$ \Call{ComputeCellId}{$p$}
%			\State $newId =$ cellStarts[$cell$] + (cellSizes[$cell$]++)
%			\State reorderedParticles[$newId$] $= p$
%		\EndFor
%		\State \Return reorderedParticles, cellStarts, cellSizes
%		\EndProcedure
%	\end{algorithmic}
%\end{algorithm}

The force evaluation is typically the most time-consuming operation of each time-step.
We map each particle to a \GPU thread which scans the adjacent cells and calculates all the interactions for the given particle.
The ordering of the particles in memory due to cell-lists increases data locality which accelerates the fetching of the particle data through cache.
We observe that exploiting the symmetry of the forces yields in faster execution despite the additional atomic operations.

\subsection{Particle bounce-back}

An important part of a microfluidics simulation is to maintain no-through properties of the particles with respect to rigid bodies, walls and membranes.
In all the three cases we introduce a continuous ``inside-outside'' function of the particle coordinates that changes its sign on the impenetrable boundary.
For example, for the wall that function is the \SDF. % SDFfigures. -> , for a sphere it is the particle distance from the sphere surface, etc.
By equating the ``inside-outside'' function with zero we obtain an equation whose solution  gives the exact collision location.
After this location  is found, we place the particle into the collision point and reverse its velocity in the frame of reference of the surface.
In order to reduce the computational cost, we exploit the cell-list that is built in the beginning of each time-step and only check the particles that are located in the cells close to the zero level of the ``inside-outside'' function.

\subsection{Efficient parallelization: compute/IO overlap}
\label{sec:impl_async}
The vastly different scales of bandwidth provided by the \GPU, PCI-E bus and the HDD storage make it necessary to overlap the intensive computation with different I/O operations performed by the code, such as MPI communications and dumping data on the disk~\cite{pacheco2011introduction}.
This is achieved by the two layers in the \mir design.

First, we run 2 MPI tasks for every computational subdomain. 
One of the tasks, called \textit{compute} task, performs the actual time-stepping on the \GPU, while the other one (\textit{postprocess} task) is responsible for all the heavy I/O and in-situ data post-processing on the CPU.
Such asynchronous design ensures perfect overlap of the disk operations with the simulation, improves code modularity and adds flexibility, since heavy data processing can be performed in parallel to the simulation, on the otherwise idling CPU.
\cref{fig:bench_datadump} shows the importance of the asynchronous HDF5 writes for the overall execution time.

The second layer of overlapping operations with each other is utilized to hide the MPI and PCI-E latencies.
Since the total number of fine-grained tasks in the time-step pipeline is about 30, maintaining their dependencies and concurrently executing some of the kernels becomes a tedious task. 
To facilitate the setup, we have implemented a \GPU-aware task scheduler based on the Kahn's topological sorting algorithm~\cite{Kahn1962}, that supports task execution on concurrent CUDA streams.
%A schematic of the task graph for a single time-step is shown if \cref{fig:tasks}, that exposes its complexity and amount of potential overlap, that can hardly be exploited by manual scheduling. 
With the help of the scheduler, we can easily overlap the halo host-to-device and device-to-host memory transfers together and the corresponding MPI communications with the force computations and potentially other heavy kernels.
%See \cref{sec:bench_single} for performance evaluations of the approach.

%\begin{figure*}
%	\caption{Dependency graph of various separate tasks being executed during one time-step. Red is the first task for the time-step, blue is the last one.}
%	\label{fig:tasks}
%	\centering
%    \includegraphics[width=0.9\textwidth]{task_graph_vert.pdf}
%\end{figure*}

\subsection{Python interface}

As software complexity increases to address multiple setups, the complexity of its usage is increasing accordingly.
Software packages often provide custom syntax for their configuration files, or even introduce a simple programming language to help users~\cite{Plimpton1995,Berendsen1995}.
We believe that implementing simulation setup through a well established programming language is superior with respect to the software specific approaches, as it benefits from the mature infrastructure and widespread usage of the language.
With its flexibility and extensive support for scientific computations via comprehensive numerical libraries, Python proves to be one the best front-end languages~\cite{Tosenberger2011} for complex codes, such as \mir.
The \textit{pybind11} project~\cite{pybind11} allowed us to easily provide a C++/CUDA proxy into Python with minimal coding efforts.

We expose all our data holder classes, handlers, plugins and the coordinator class such that the user is able to assemble the specific simulation setup out of the few basic building blocks like a construction toy.
Further advances of our approach include a very thin abstraction layer and the ease of documenting the functions available to the end users.

\section{Validation}
\label{sec:validation}

In this section we present a set of validation cases for  \mir, in which we compare our results against available analytical solutions or previously published data.
An additional large set of more fine-grained tests (for example, bounce-back tests, momentum conservation verification, etc.) is available with the source code. We note that \mir was developed using experiences from a previous code (\udx~\cite{Rossinelli2015}) co-developed by our group. These experiences were instrumental in developing a code that is shown to outperform \udx, a Gordon bell finalist in 2015~\cite{Rossinelli2015}.

\subsection{Viscosity of the \DPD liquid for different parameters}

\Cref{tab:viscosity} shows the measured viscosity of the \DPD fluid with mass density $\rho$ given specific parameters compared to the one presented in the literature.
We chose to measure the viscosity using a Poiseuille flow, such that $\eta = \rho f R^2 / (8 u_{avg})$.
Here $R$ is the radius of the pipe, $f$ is the body force applied on each particle in order to form the pressure gradient, and $u_{avg}$ is the average flow velocity.
We assume small enough time-step where in case it is not reported, and obtain the range of viscosities for body force ranging from $0.0003$ to $0.05$, obtaining good agreement with the previously reported values.

\begin{table*}[ht]
	\caption{Comparison of the \DPD fluid viscosity obtained from different parameters available in the literature with \mir.
	We simulate a Poiseuille flow inside a circular pipe of radius $R = 20$, length $L=3R=60$, and ran $10^6$ time-steps.
	The pressure gradient was applied by adding body-force on each particle ranging from $0.0003$ to $0.05$.
	The viscosity in obtained by averaging the velocity over the last $2\times10^5$ steps. }
	\label{tab:viscosity}
	\vspace{5pt}
	\centering
	%\small
	\begin{tabular}{@{}lllllllcrr@{}}
		\multicolumn{7}{c}{\DPD parameters} & \multirow{2}{*}{$\eta_{\text{\mir}}$} & \multirow{2}{*}{$\eta_{\text{ref}}$} & \multirow{2}{*}{Reference} \\
		%\cmidrule{1-7}
		$a$ & $\gamma$ & $\rho$ & $k$ & $k_B T$ & $r_c$ & $\Delta t$ & & & \\
		\toprule
		0.9375 &      115.6 &          4 &          1 &       0.05 &          1 &       0.01 &   $\numrange{4.4}{4.6}$     &   4.7    & \cite{Li2008}      \\
	         6 &         20 &          3 &       0.15 &        0.1 &          1 &      0.002 &   $\numrange{8.0}{8.3}$     &   8.1    & \cite{Fedosov2011} \\
	         4 &          8 &          3 &       0.15 &        0.1 &        1.5 &      0.001 &   $\numrange{24.7}{26.3}$   &   26.3   & \cite{Fedosov2011} \\
	         4 &         40 &          3 &       0.15 &        0.1 &        1.5 &     0.0002 &   $\numrange{122.5}{129.5}$ &   126    & \cite{Fedosov2011} \\
	        25 &       6.75 &          3 &          1 &          1 &          1 &       0.04 &   $\numrange{0.89}{0.9}$    &   0.91   & \cite{Groot1997}   \\
	     18.75 &        4.5 &          4 &          1 &          1 &          1 &      0.005 &   $\numrange{1.07}{1.08}$   &   1.08   & \cite{Fan2006}    \\
	     18.75 &        4.5 &          4 &       0.25 &          1 &          1 &      0.005 &   $\numrange{2.44}{2.45}$   &   2.59   & \cite{Fan2006}    \\
	         0 &      20.25 &          6 &          1 &        0.5 &          1 &       0.01 &   $\numrange{2.08}{2.1}$    &   2.09   & \cite{Backer2005}
	\end{tabular}
\end{table*}

\subsection{Periodic Poiseuille flow}
\label{sec:periodic_pois}

\begin{figure}
  \caption {
    Velocity profile in a periodic Poiseuille setup from simulations (symbols) and analytical solution (solid line). $L=64$, $\rho=8$, $a=10$, $\gamma=20$, $k_B T=1.0$, $k=0.5$, $\Delta t = 0.005$.
  }
  \label{fig:val_per_pois}
  \centering
  \includegraphics{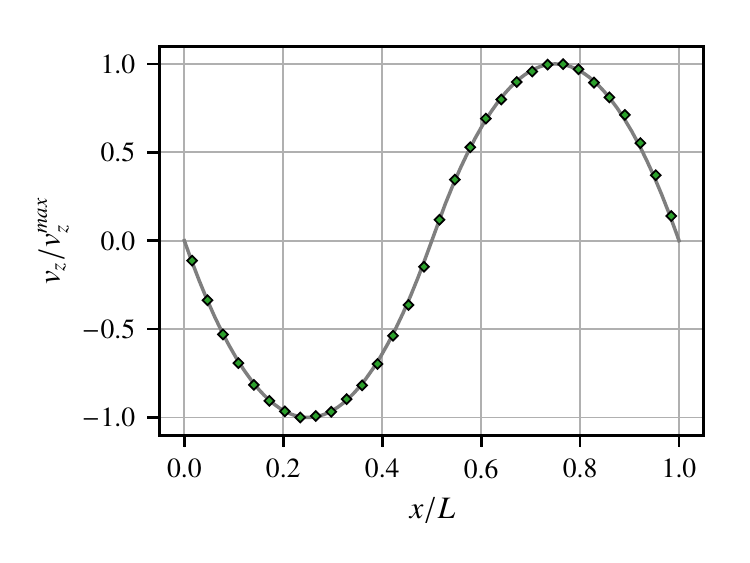}
\end{figure}

Periodic Poiseuille flow was introduced in~\cite{Backer2005} as a convenient way to measure viscosity of a particle fluid without walls.
The setup consists of a cubic domain ($L \times L \times L$) with periodic boundary conditions and the space-dependent body force that drives the fluid in the opposite directions:
\begin{equation}
  \mathbf{f}(\mathbf{r}) = \begin{cases} (0, 0, -f), & r_x \leqslant L/2,\\ (0,0,f), & r_x > L/2. \end{cases}
\end{equation}
For a Newtonian fluid, the resulting laminar flow has a parabolic profile: $v_z(x) = \rho f (x L/2 - x^2) / 2 \eta$, where $\rho$ is the fluid's mass density and $\eta$ its dynamic viscosity.
The simulation results are depicted in \cref{fig:val_per_pois}.

\subsection{Taylor-Couette flow}

\begin{figure}
  \caption {
    Fluid velocity in the azimuthal direction against radial coordinate from simulation (symbols) and analytical solution (solid line), obtained with $R_{in}=10$, $R_{out}=32$, $\omega_{in}=0$, $\omega_{iout}=0.01$, $\rho=10$, $a=10$, $\gamma=10$, $k_B T=0.5$, $k=0.125$, $\Delta t = 0.001$
  }
  \label{fig:val_taylor_couette}
  \centering
  \includegraphics{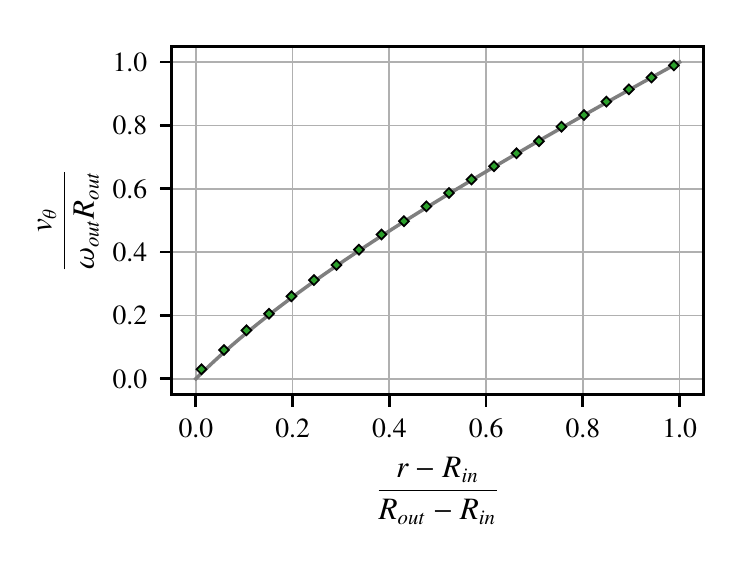}
\end{figure}

Taylor-Couette flow consists of a fluid moving between two concentric cylinders, one of them rotating with respect to the other.
Given the cylinders radii $R_{in}$ and $R_{out}$, and their rotational velocities $\omega_{in}$ and $\omega_{out}$, the resulting azimuthal velocity of the fluid is given by the following:
\begin{equation}
  v(r) = r \, \omega_{in} \frac{\mu - \eta^2}{1 - \eta^2} + \frac{1}{r} \,\omega_{in} \, R_{in}^2 \frac{1 - \mu}{1 - \eta^2}, \\
\end{equation}
where $\mu =  \omega_{out} / \omega_{in}$ and $\eta = R_{in} / R_{out}$.
The simulation results are depicted in \cref{fig:val_taylor_couette}.

\subsection{Jeffery orbits in shear flow}

\begin{figure}
  \caption{Evolution of the inclination angle $\phi$ of the longer ellipsoid axis with respect to the flow direction with $a = 5$, $b = 3$, $\Gamma = 0.1$:  Simulation (symbols) and Jeffrey's theory (solid line). \DPD parameters: $L=64$, $\rho=8$, $a=25$, $\gamma=50$, $k_B T=0.5$, $k=0.5$, $\Delta t = 0.005$.}
	\label{fig:val_jeffry}
	\centering
    \includegraphics{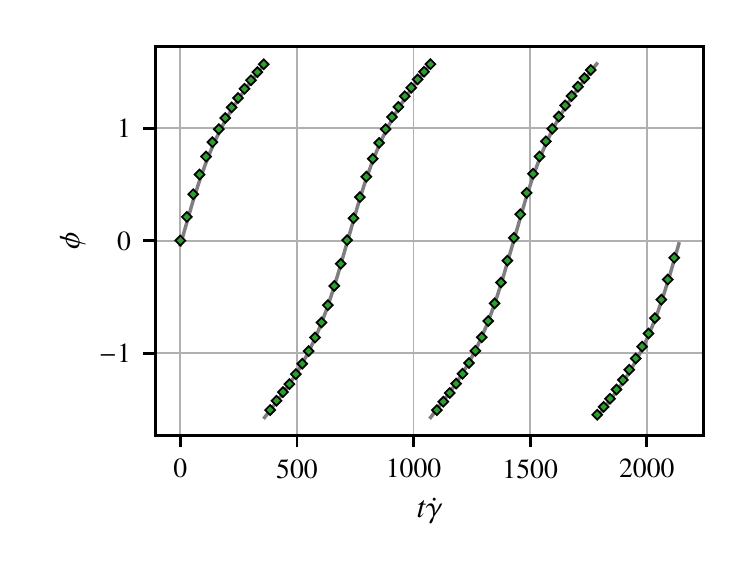}
\end{figure}

We validate the rigid body dynamics by simulating the rotation of an ellipsoid in a simple shear flow.
In the limit of small Reynolds number, the inclination angle of the longer ellipsoid axis is known to be following the Jeffery orbit~\cite{Jeffery1922} over time:
\begin{equation}  %np.arctan( b/a * np.tan(a*b*G * t / (a**2 + b**2)) )
	\phi(t) = \arctan\left(  \frac{b}{a} \, \tan{\frac{a b \, \dot{\gamma} t}{a^2 + b^2}}  \right),
\end{equation}
where $a$ and $b$ is the longer and shorter axes of the ellipsoid and $\dot{\gamma}$ is the shear rate.
We enforce the shear profile by moving two parallel plates with opposite velocities.
The ellipsoid is kept in the middle of the computational domain throughout the entire simulation.
%We use two moving walls to enforce shear flow and keep the ellipsoid in the middle of the computational domain.
The results are depicted in the \cref{fig:val_jeffry}.

\subsection{Flow past a sphere close to a wall}

\begin{figure}
	\caption{Drag and lift coefficient for a sphere translating in the quiescent fluid at a distance $L = 2r$ from the infinite wall. Error bars represent 2 standard deviations of the mean estimate. \DPD parameters vary to satisfy given $\Rey$ and low enough Mach number $\mathrm{Ma} < 0.2$.  }
	\label{fig:val_sphere_lift}
	\centering
    \includegraphics{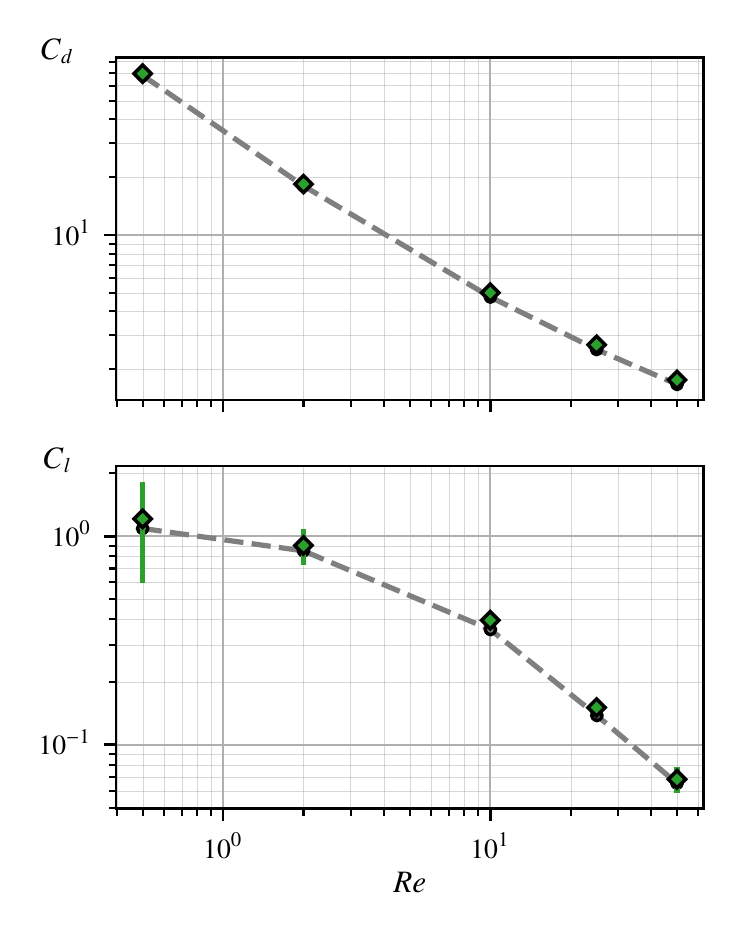}
\end{figure}

We study the drag and lift coefficients of a sphere translating in the otherwise quiescent fluid close to a flat wall.
The setup is characterized (see \cref{fig:val_sphere_lift}) by the fluid kinematic viscosity $\nu$, sphere radius $r$, wall distance $L$ and the translation velocity $u$.
We perform the simulation in the sphere frame of reference by keeping the sphere center of mass forcedly fixed.
We use periodic boundary conditions in $x$ and $y$ directions and introduce two walls orthogonal to the $z$ direction.
The walls are moved with the velocity $-u$ and the average fluid velocity in the domain is kept constant.
The remaining non-uniform wake downstream the sphere is removed in a thin layer before the domain boundary.
The sphere, however, is allowed to rotate freely.

The quantities of interest are the lift and drag coefficients of the sphere, which are computed by time-averaging of the fluid forces acting on the sphere.
Due to the third Newton's law those forces are simply negation of the forces that are required to keep the sphere in place.
The non-dimensional expression of the above quantities read:
\begin{equation}
	C_{l,d} = \frac{\left< F_{l,d} \right>}{\frac{1}{2} \rho u^2 \, \pi r^2 },
\end{equation}
where the angular brackets denote the time-averaging and the subscripts $l,d$ represent wall-normal lift and wall-parallel drag, respectively.
The simulation results are depicted in \cref{fig:val_sphere_lift}.
Here we consider the case with $L=2r$ and vary the Reynolds number $\Rey = L u / \nu$ from $0.5$ to $50$.

We obtain a good correspondence against the previous simulations carried out with spectral methods~\cite{ZENG2005}.
A noticeable discrepancy in for $\Rey = 10,20$ can attributed to the fact that we used a little smaller domain size in order to reduce the computational cost.

\subsection{Cell stretching}
\begin{figure}
  \caption {
    Axial and transverse diameters of the \RBC against stretching force.
    The simulation (symbols) employs parameters corresponding to the best fit~\cite{Fedosov2010} from experimental data (solid line).
  }
  \label{fig:val_stretch}
  \centering
  \includegraphics{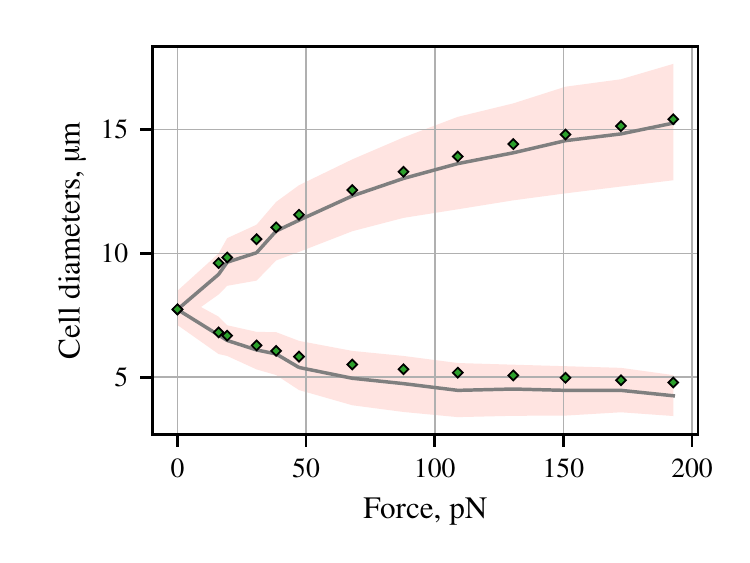}
\end{figure}

%For the cell model validation we employ the same experiment used in~\cite{Fedosov2010}: the cell stretching with optical tweezers.
We validate the \RBC model described in \cref{sec:model:objects} by simulating a \RBC stretched by optical tweezers and comparing the force-extension curve with the experimental data~\cite{Fedosov2010}.
We vary the force applied to the few opposite particles of a single cells membrane, and measure the axial and transverse diameters of the cell.
The results are depicted in the \cref{fig:val_stretch}.

 \section{Performance}
\label{sec:benchmarks}

One of the advantages of the \mir software is the very fast time to solution and nearly perfect weak scaling up to hundreds of nodes.
We benchmark our software against the state-of-the-art \MD packages (HOOMD-Blue and LAMMPS) and \udx code~\cite{Rossinelli2015} running on \GPU clusters, and perform strong and weak scaling studies.
All the timings were collected with the standard \verb|nvprof| CUDA profiler and the internal high-resolution system clock.
We used several hardware platforms to obtain the results: Piz Daint supercomputer (CSCS, Switzerland) with one \nv Tesla P100 per node, Leonhard cluster (ETHZ, Switzerland) with 8 \nv GTX 1080Ti per node, Microsoft Azure platform with \nv Tesla V100 and a high-end consumer laptop with \nv GTX 1070.

\subsection{Periodic Poiseuille flow}
\label{sec:pp_bench}
Our first benchmark is the periodic Poiseuille flow of the \DPD particles, the least complex setup that is nevertheless representative for a wide class of problems where the object of cell suspension is dilute.
We consider a cubic domain of size $L \times L \times L$ per every \GPU, filled uniformly with the \DPD particles at a constant density $\rho$ with the periodic body force $f$ (see \cref{sec:periodic_pois}).
%First, we report time distribution for various parts of the code with completely synchronous \GPU kernel execution (forced by setting \verb|CUDA_LAUNCH_BLOCKING| environmental variable).
%The reason for using synchronous timings is that in the \textit{production} asynchronous scenario several independent computational kernels may overlap, thus sharing the \GPU and yielding in longer individual execution times.
%So in order to accurately estimate the performance of each kernel, we use the synchronous mode, while to obtain the overall wall-clock time per simulation step we use the faster asynchronous mode. 

The reference benchmark employs $L = 64$ and $\rho = 8$, resulting in total of $2.1$M particles, or $12.6$M degrees of freedom, and roughly $34.1$M interacting pairs per node assuming uniform particle distribution.
%From the \Cref{fig:kernels_dpd} it is clear that the force calculation takes the dominant part of the run-time, taking as much as 87\% of the total runtime.
The average time-step on 1 compute node of Piz Daint is $7.01$ms, which results in throughput of $4.9$ billion interactions per second per \GPU node.

\cref{tab:compare_dpd} summarizes the performance comparison against \udx code and HOOMD-blue on the Piz Daint supercomputer. Since the choice of the simulation parameters may affect the run-time, in \cref{tab:compare_vs_hoomd} we show that our code consistently outperforms HOOMD-blue for various benchmark setups.

\begin{table}
	\caption{Wall-clock time in ms per one simulation time-step on the Piz Daint supercomputer (\nv P100 \GPUs). $2^{21}$ particles per node, \DPD parameters are the following: $a=50$, $k_bT = 1$, $\gamma = 20$, $\delta t = 0.002$. The neighbor-list parameters of HOOMD-blue are tuned for the best performance \label{tab:compare_dpd}}
	\vspace{10pt}
	\centering
	\begin{tabular}{@{}llll@{}}
	     &  \multicolumn{3}{c}{Nodes} \\
		 \cmidrule{2-4}
		 & 1 & 27 & 64 \\
		\toprule
		HOOMD-blue & 11.1 & 18.0 & 18.3 \\
		\udx        & 10.4 & 11.3 & 11.6 \\
		\mir      & \textbf{7.0} & \textbf{7.3} & \textbf{7.3}
	\end{tabular}
\end{table}

% vs HOOMD
%0.05_0.02:  1.554123 - 2.190813,  1.804310
%0.5_0.02:  2.762361 - 7.816051,  4.195252
%5_0.02:  7.200886 - 22.005712,  12.561225
%
%0.05_0.005:  1.010952 - 1.393937,  1.212886
%0.5_0.005:  1.384808 - 2.276542,  1.783543
%5_0.005:  2.647407 - 5.435112,  3.877915
%
%0.05_0.002:  0.898159 - 1.196880,  1.051118
%0.5_0.002:  1.133326 - 1.632636,  1.353550
%5_0.002:  1.620205 - 2.834074,  2.146691

\begin{table}
	\caption{Performance comparison against HOOMD-Blue. Wall-clock time in ms per one simulation time-step on the Piz Daint supercomputer (\nv P100 \GPUs). Domain size is always $64^3$, \DPD parameters vary: $a = 10$, $\gamma \in \{1, 10, 100\}$, $\rho \in \{4,8,12\}$, $r_c \in \{0.8, 1.0, 1.2\}$. The neighbor-list parameters of HOOMD-blue are tuned for the best performance. Reported speedup is $t_{HOOMD} / t_{\text{\mir}}$. \label{tab:compare_vs_hoomd} }
	\vspace{10pt}
	\centering
	\begin{tabular}{@{}llcc@{}}
		$k_B T$ & $\Delta t$ & Speedup range & Average speedup \\
		\toprule
		0.05 & 0.02  & 1.6 -- 2.2  & 1.8  \\
		0.5  & 0.02  & 2.8 -- 7.8  & 4.2  \\
		5.0  & 0.02  & \,\,\,7.2 -- 22.0 & 12.6 \\
		0.05 & 0.005 & 1.0 -- 1.4  & 1.2  \\
		0.5  & 0.005 & 1.4 -- 2.3  & 1.8  \\
		5.0  & 0.005 & 2.6 -- 5.4  & 3.9  \\
		0.05 & 0.002 & 0.9 -- 1.2  & 1.1  \\
		0.5  & 0.002 & 1.1 -- 1.6  & 1.4  \\
		5.0  & 0.002 & 1.6 -- 2.8  & 2.1  \\
	\end{tabular}
\end{table}

\begin{figure}[ht]
 	\caption { Weak (top) and strong (bottom) scaling efficiency of periodic Poiseuille benchmark for different subdomain size. Particle density $\rho = 8$ for all the runs.  }
    \centering
    \includegraphics{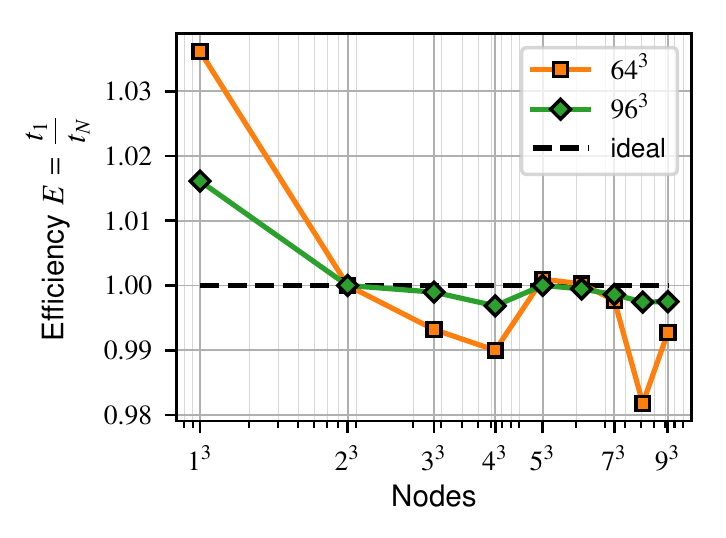}
    \includegraphics{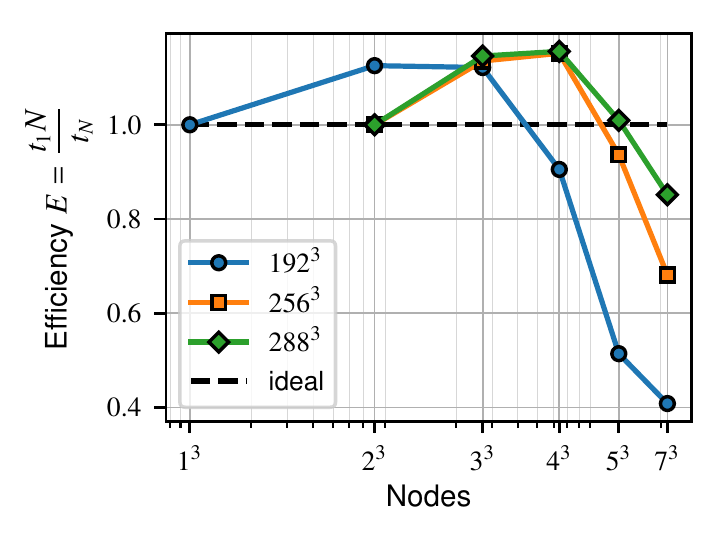}
	\label{fig:bench_poiseuille}
\end{figure}

\Cref{fig:bench_poiseuille}~(top) shows weak scaling capabilities of \mir running periodic Poiseuille benchmark on Piz Daint.
Note that the reference point was chosen at $N = 8$ nodes, as the single node execution employs some optimizations eliminating almost all of the MPI communication. 
Due to the good compute/transfer overlap, we reach almost perfect weak scaling for up to $1000$ nodes.
Strong scaling is not the primary scope of our code, as typically the problems of interest consist of very many particles.
However, the strong efficiency of \mir is still good, see \Cref{fig:bench_poiseuille}~(bottom).
It also shows super-linear behavior that we believe is attributed to better spatial locality of smaller amount of data in the cache, which is the main bottleneck in computing interactions.

\begin{figure}[ht]
 	\caption { Periodic Poiseuille benchmark on Piz Daint with data dumps every $100$ steps.
}
    \centering
    \includegraphics{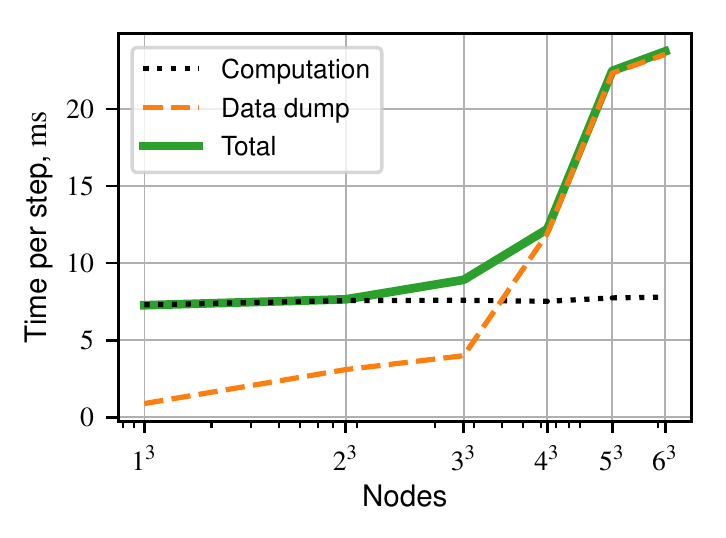}
	\label{fig:bench_datadump}
\end{figure}

\Cref{fig:bench_datadump} shows benefits of the overlapping computations with I/O.
We ran the periodic Poiseuille benchmark on Piz Daint with dumping HDF5 flow fields every $100$ steps.
The I/O bandwidth doesn't scale well with the number of nodes, reaching about \SI{3}{\giga\byte\per\second} for ${\sim}64$ nodes.
As one can observe, data dumps are overlapped with the computations, making the total runtime approximately maximum of the I/O and calculations. Note that in typical simulations, the dump frequency is much lower, such that we are never limited by the I/O performance.

\subsection{Periodic whole blood flow}
The second representative benchmark is the periodic Poiseuille flow of the blood, which include cell membranes, different viscosities between the plasma and the cytoplasm and runs with activated bounce-back mechanism.
The latter incurs the biggest performance penalty, because it requires that objects are exchanged over MPI \textit{after} the integration but \textit{before} the bounce-back itself is performed.
Therefore possibilities to overlap communication of objects and computation are very limited in this case.

\begin{table}[ht]
	\caption{Wall-clock time in ms per one simulation time-step. Periodic whole blood at $35\%$ hematocrit level, $1.6M$ particles per node. }
	\label{tab:compare_blood}
	\vspace{5pt}
	\centering
	\begin{tabular}{@{}llll@{}}
	     &  \multicolumn{3}{c}{Nodes} \\
		 \cmidrule{2-4}
		 & 1 & 8 & 27\\
		\toprule
		LAMMPS USER-MESO 2.0 &        140.2 &         144.1 &        143.8 \\
		\mir                 & \textbf{9.8} & \textbf{13.6} & \textbf{13.7}
	\end{tabular}
\end{table}

We consider a cubic domain of size $L \times L \times L$ filled uniformly with the \RBCs at a specific volume fraction (or \textit{hematocrit} level) $Ht$.
The fluid density is $\rho$ and the periodic force $f$ is applied in the same manner as for the previous case.
%The \Cref{fig:kernels_blood} depicts the time distribution for various parts of the code again in the two variants: synchronous and asynchronous.
%All the timings were collected on the Piz Daint supercomputer with the standard \verb|nvprof| CUDA profiler.

\Cref{tab:compare_blood} summarizes the performance comparison against the only \GPU code known to the authors with roughly similar feature set: LAMMPS USER-MESO $2.0$~\cite{Blumers2017}.
We used the set-up available with the USER-MESO that runs whole blood at $Ht = 35\%$ and set the domain to roughly $76 \times 58 \times 58$.
The present implementation outperforms USER-MESO by a factor of ${\sim}14$x (${\sim}11$x on many nodes), mainly attributed to the fact that LAMMPS, although evaluating forces on the \GPU, keeps and uses a lot of supporting data on the CPU, incurring slow PCI-E traffic.

\begin{figure}[ht]
 	\caption { Weak (top) and strong (bottom) scaling efficiency of periodic blood benchmark for different domain sizes. $\rho = 8$, $Ht = 40\%$. }
    \centering
    \includegraphics{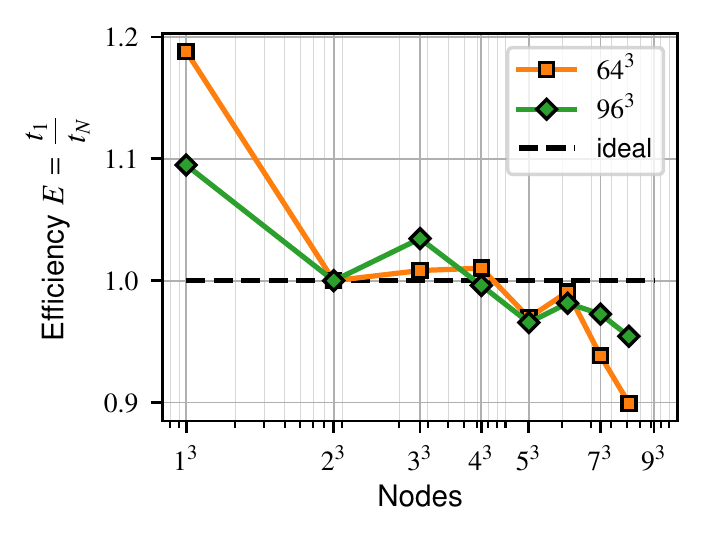}
    \includegraphics{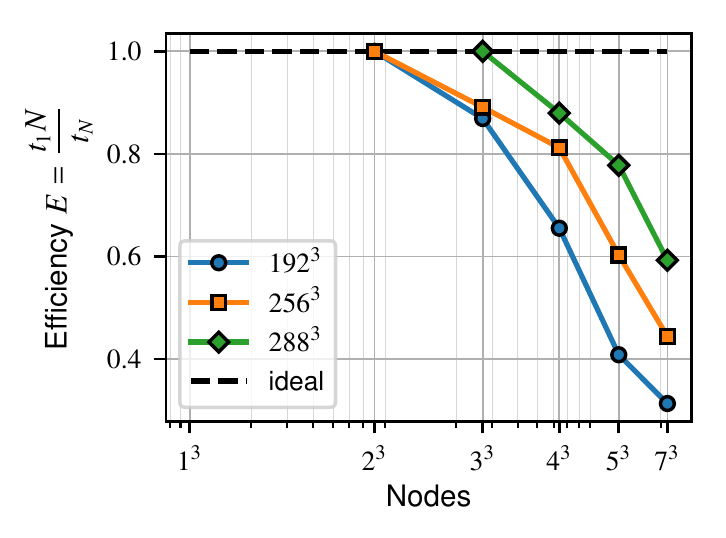}
	\label{fig:bench_blood}
\end{figure}

\Cref{fig:bench_blood}~(top) shows weak scaling capabilities of \mir running periodic blood benchmark at $Ht = 40\%$.
The compute/transfer overlap worsens compared to the pure liquid flow, resulting in unstable execution time and deteriorated scaling.
For the same reason we observe that the single node case benefits significantly more from the MPI calls elimination.
Nevertheless, the code reaches $95\%$ efficiency on $512$ nodes for a bigger subdomain size.
Strong scaling is also worse compared to the simple \DPD case, but still yields in about $50\%$ efficiency going from $8$ to $216$ nodes on a $256^3$ domain size, see \Cref{fig:bench_blood}~(bottom).

\subsection{Microfluidic device}
\label{sec:dld_bench}

\begin{figure}[ht]
	\caption{ Snapshot of the two-post periodic simulation (top) of a \DLD device with irregularly-shaped obstacles and time distribution
	per kernel (bottom) on a single Piz Daint node.
	The domain size is $64\times56\times60$, particle density $\rho=8$, hematocrit $Ht = 40\%$.}
	\label{fig:dld_bench}
	\centering
	\includegraphics[width=0.99\columnwidth]{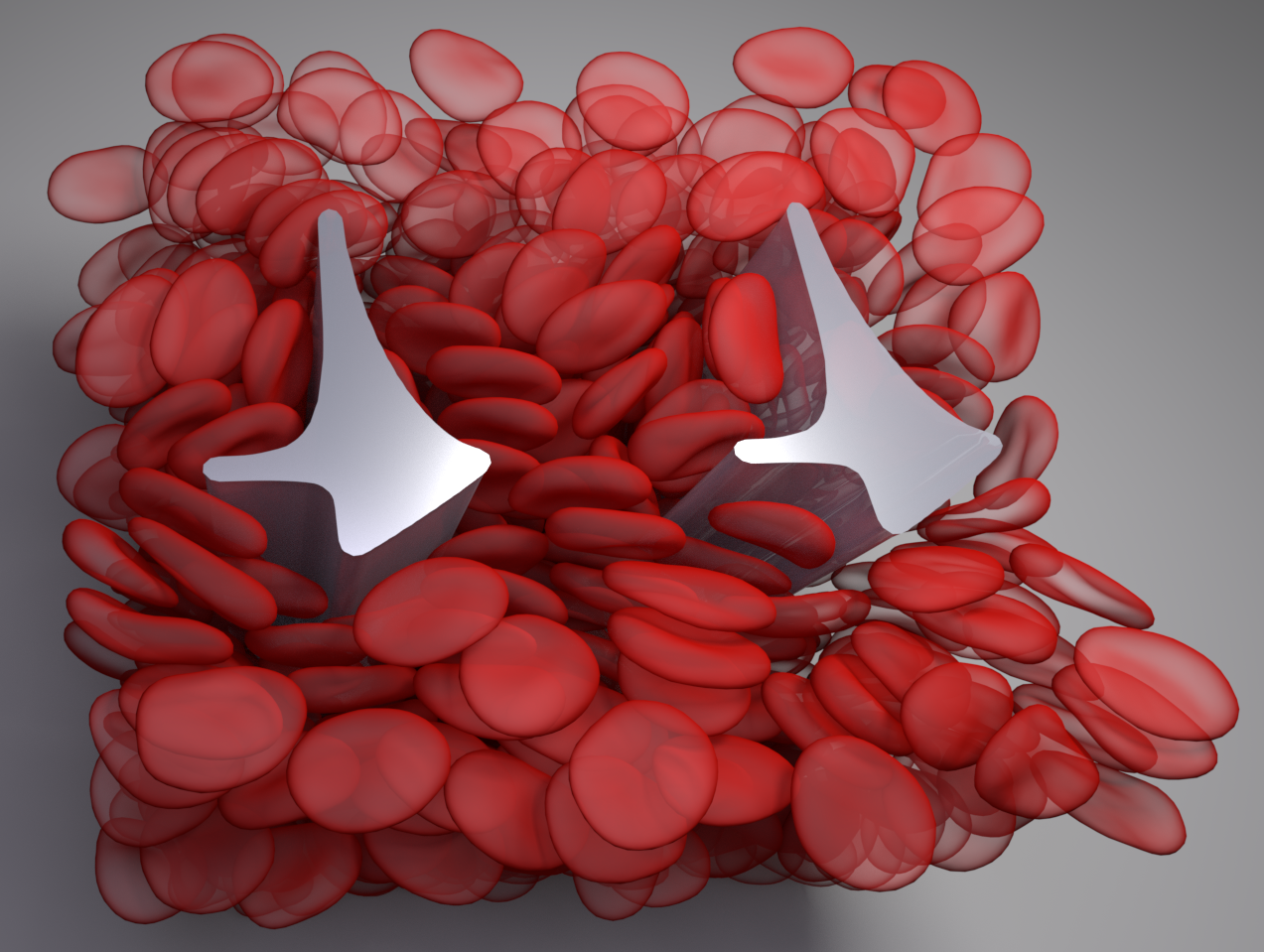}\\
	\vspace{10pt}
	\includegraphics[width=0.8\columnwidth]{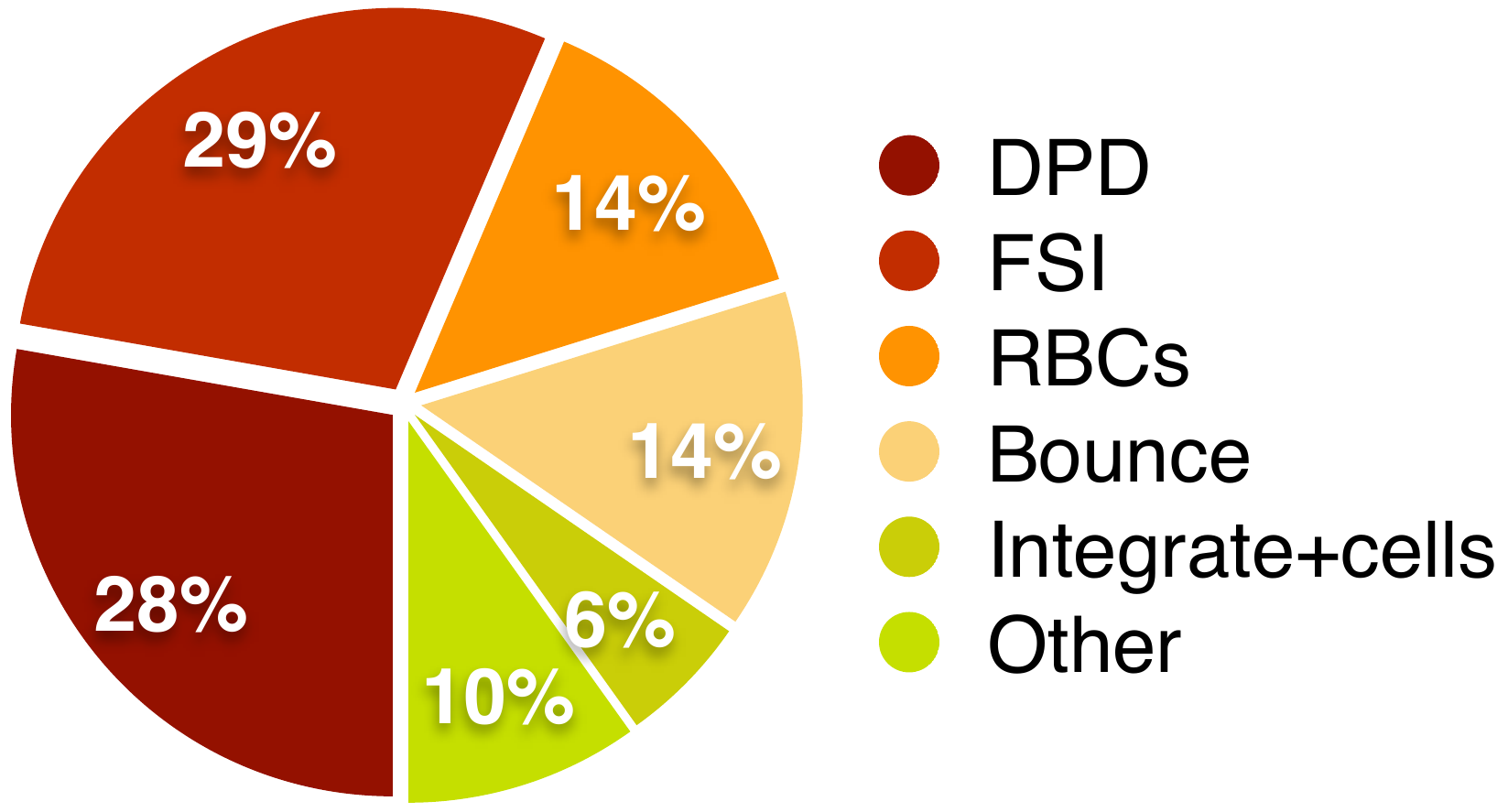}
\end{figure}

To assess \mir at full complexity we simulate a part of a microfluidic device that captures \CTCs from the whole blood~\cite{Karabacak2014a}.
The first stage of the device exploits \DLD principle~\cite{Huang2004a} to separate the bigger and stiffer \CTCs from the smaller and very flexible \RBCs.
%The idea is that as the flow passes a tilted array of obstacles (posts), the \RBCs follow the streamlines and freely go through them, while the \CTCs are deflected from the obstacles and get displaced with respect to the average flow direction.
In order to study the device and, later, to optimize the flow parameters and shape of the obstacles, we model its small part with two posts and impose periodic boundary conditions such that the domain replications correspond to the full device, see \Cref{fig:dld_bench} (top).
The setup features domain of complex shape, blood cells at $Ht = 40\%$ with cytoplasm $5$ times more viscous compared to the solvent, and all the bounce-back mechanisms to prevent particle leakage.

Here, we report time distribution for various parts of the code with completely synchronous \GPU kernel execution (forced by setting \verb|CUDA_LAUNCH_BLOCKING| environmental variable).
The reason for using synchronous timings is that in the \textit{production mode}, several independent computational kernels may overlap, thus sharing the \GPU and yielding in longer individual execution times.
So in order to accurately estimate the performance of each kernel, we use the synchronous mode, while to obtain the overall wall-clock time per simulation step we use the faster asynchronous one.
The time chart in \Cref{fig:dld_bench} (bottom) shows that fluid forces account for biggest part of the execution time: 57\%.
Bounce-back and the internal membrane forces have an equal share of 14\% each, while 6\% of the time is taken by memory-intensive integration and cell-list creating.
Remaining 10\% of the time is spent in various helper kernels, like packing/unpacking, various sorts, scans, etc.
With the help of the profiler we identify that the main bottleneck for the force kernels is the L1 and L2 cache performance, with bulk and FSI kernels reaching around 60\% of the aggregate cache bandwidth.

\subsection{Hardware comparison}
As a last step of our performance analysis, we run the periodic Poiseuille (see \cref{sec:pp_bench}) the \DLD benchmark (see \cref{sec:dld_bench}) on different available hardware platforms: a consumer-grade laptop with \nv GTX 1070, ETHZ Leonhard cluster with \nv GTX 1080Ti, Piz Daint supercomputer with \nv Tesla P100 and Microsoft Azure virtual machine with \nv Tesla V100.
\cref{tab:compare_platforms} summarizes the results.
Together with lower-level kernel analysis, they show that \mir performance benefits greatly from the better and newer hardware, with the most important factor being the size and the speed of the L1 and L2 \GPU caches.

\begin{table}[ht]
	\caption{Wall-clock time in \si{\milli\second} per one simulation time-step on different \GPUs.
	All runs are single-node.
	PP means periodic Poiseuille benchmark, see \cref{sec:pp_bench}, DLD means microfluidic device, see \cref{sec:dld_bench}. 	\label{tab:compare_platforms}}
	\vspace{10pt}
	\centering
	\begin{tabular}{@{}lll@{}}
	      & PP & DLD \\
		\toprule
		1070M & 12.6 & 27.8 \\
		1080Ti & 6.9 & 18.8 \\
		P100 & 7.0 & 15.9 \\
		V100 & 3.7 & 8.8 \\
	\end{tabular}
\end{table}

 \section{Summary}
\label{sec:conclusion}

In this paper we introduced the open-source \GPU software package \mir, that implements the \DPD method.
Our code can handle arbitrarily complex domains, many visco-elastic \RBCs and rigid bodies of various shapes, parallelized over hundreds of \GPU nodes.
Such set of features, up to our knowledge, is not offered by any other particle-based simulation software like \cite{Rossinelli2015,Blumers2017}.
We also presented a set of validation cases that exhibit good correspondence of simulation results with analytical, experimental or earlier numerical data.

With extensive benchmarking, we showed that \mir provides very fast time-to-solution, efficiently harnessing \GPU capability.
Our code outperforms the state-of-the-art competitors by factors of ${\sim}1.5$ (for pure \DPD liquid) up to $11$ (for dense blood) and reaches high weak and strong scaling efficiencies for up to 512 nodes of Piz Daint supercomputer.
Furthermore, \mir comes with the extensively documented Python interface, that offers a simple mechanism to combine the implemented features into a complex simulation. 
The code is distributed as open-source at \href{https://github.com/cselab/Mirheo}{https://github.com/cselab/Mirheo}.

%% Appendix may be the user docs from the readthedocs, or maybe not



%% References


\bibliographystyle{elsarticle-num}
\bibliography{bibliography.bib}

\begin{thebibliography}{10}
\expandafter\ifx\csname url\endcsname\relax
  \def\url#1{\texttt{#1}}\fi
\expandafter\ifx\csname urlprefix\endcsname\relax\def\urlprefix{URL }\fi
\expandafter\ifx\csname href\endcsname\relax
  \def\href#1#2{#2} \def\path#1{#1}\fi

\bibitem{Beebe2002}
D.~J. Beebe, G.~A. Mensing, G.~M. Walker,
  \href{http://www.annualreviews.org/doi/10.1146/annurev.bioeng.4.112601.125916}{{Physics
  and Applications of Microfluidics in Biology}}, Annual Review of Biomedical
  Engineering 4~(1) (2002) 261--286.
\newblock \href {https://doi.org/10.1146/annurev.bioeng.4.112601.125916}
  {\path{doi:10.1146/annurev.bioeng.4.112601.125916}}.
\newline\urlprefix\url{http://www.annualreviews.org/doi/10.1146/annurev.bioeng.4.112601.125916}

\bibitem{Whitesides2006}
G.~M. Whitesides, {The origins and the future of microfluidics}, Nature
  442~(7101) (2006) 368--373.
\newblock \href {http://arxiv.org/abs/arXiv:1011.1669v3}
  {\path{arXiv:arXiv:1011.1669v3}}, \href
  {https://doi.org/10.1016/j.agee.2012.07.026}
  {\path{doi:10.1016/j.agee.2012.07.026}}.

\bibitem{Streets2013}
A.~M. Streets, Y.~Huang, {Chip in a lab: Microfluidics for next generation life
  science research}, Biomicrofluidics 7~(1) (2013).
\newblock \href {https://doi.org/10.1063/1.4789751}
  {\path{doi:10.1063/1.4789751}}.

\bibitem{Squires2005}
T.~M. Squires, S.~Quake, {Microfluidics: Fluid physics at the nanoliter},
  Reviews of Modern Physics 77~(July) (2005).

\bibitem{Sackmann2014}
E.~K. Sackmann, A.~L. Fulton, D.~J. Beebe,
  \href{http://dx.doi.org/10.1038/nature13118}{{The present and future role of
  microfluidics in biomedical research}}, Nature 507~(7491) (2014) 181--189.
\newblock \href {https://doi.org/10.1038/nature13118}
  {\path{doi:10.1038/nature13118}}.
\newline\urlprefix\url{http://dx.doi.org/10.1038/nature13118}

\bibitem{Zhang2016}
J.~Zhang, S.~Yan, D.~Yuan, G.~Alici, N.~T. Nguyen, M.~{Ebrahimi Warkiani},
  W.~Li, {Fundamentals and applications of inertial microfluidics: A review},
  Lab on a Chip 16~(1) (2016) 10--34.
\newblock \href {https://doi.org/10.1039/c5lc01159k}
  {\path{doi:10.1039/c5lc01159k}}.

\bibitem{Clague2001}
D.~Clague, {Computer simulations help microfluidic device designers get from
  concept to prototype quickly and efficiently.}, S{\&}Tr (2001) 4--11.

\bibitem{DiCarlo2009}
D.~{Di Carlo}, J.~F. Edd, K.~J. Humphry, H.~A. Stone, M.~Toner, {Particle
  segregation and dynamics in confined flows}, Physical Review Letters 102~(9)
  (2009) 1--4.
\newblock \href {http://arxiv.org/abs/NIHMS150003} {\path{arXiv:NIHMS150003}},
  \href {https://doi.org/10.1103/PhysRevLett.102.094503}
  {\path{doi:10.1103/PhysRevLett.102.094503}}.

\bibitem{Cimrak2012}
I.~Cimr{\'{a}}k, M.~Gusenbauer, T.~Schrefl,
  \href{http://dx.doi.org/10.1016/j.camwa.2012.01.062}{{Modelling and
  simulation of processes in microfluidic devices for biomedical
  applications}}, Computers and Mathematics with Applications 64~(3) (2012)
  278--288.
\newblock \href {https://doi.org/10.1016/j.camwa.2012.01.062}
  {\path{doi:10.1016/j.camwa.2012.01.062}}.
\newline\urlprefix\url{http://dx.doi.org/10.1016/j.camwa.2012.01.062}

\bibitem{Karabacak2014}
N.~M. Karabacak, P.~S. Spuhler, F.~Fachin, E.~J. Lim, V.~Pai, E.~Ozkumur, J.~M.
  Martel, N.~Kojic, K.~Smith, P.-i. Chen, J.~Yang, H.~Hwang, B.~Morgan,
  J.~Trautwein, T.~A. Barber, S.~L. Stott, S.~Maheswaran, R.~Kapur, D.~A.
  Haber, M.~Toner, {Microfluidic, marker-free isolation of circulating tumor
  cells from blood samples}, Nature Protocols 9~(3) (2014) 694--710.
\newblock \href {https://doi.org/10.1038/nprot.2014.044}
  {\path{doi:10.1038/nprot.2014.044}}.

\bibitem{Scherr2015}
T.~Scherr, G.~L. Knapp, A.~Guitreau, D.~S.~W. Park, T.~Tiersch, K.~Nandakumar,
  W.~T. Monroe, {Microfluidics and numerical simulation as methods for
  standardization of zebrafish sperm cell activation}, Biomedical Microdevices
  17~(3) (2015).
\newblock \href {https://doi.org/10.1007/s10544-015-9957-6}
  {\path{doi:10.1007/s10544-015-9957-6}}.

\bibitem{grimmer2019}
A.~Grimmer, M.~Hamidovi{\'c}, W.~Haselmayr, R.~Wille, Advanced simulation of
  droplet microfluidics, ACM Journal on Emerging Technologies in Computing
  Systems (JETC) 15~(3) (2019) 26.

\bibitem{Freund2014}
J.~B. Freund,
  \href{http://www.annualreviews.org/doi/abs/10.1146/annurev-fluid-010313-141349}{{Numerical
  Simulation of Flowing Blood Cells}}, Annual Review of Fluid Mechanics 46~(1)
  (2014) 67--95.
\newblock \href {https://doi.org/10.1146/annurev-fluid-010313-141349}
  {\path{doi:10.1146/annurev-fluid-010313-141349}}.
\newline\urlprefix\url{http://www.annualreviews.org/doi/abs/10.1146/annurev-fluid-010313-141349}

\bibitem{rahimian2010petascale}
A.~Rahimian, I.~Lashuk, S.~Veerapaneni, A.~Chandramowlishwaran, D.~Malhotra,
  L.~Moon, R.~Sampath, A.~Shringarpure, J.~Vetter, R.~Vuduc, et~al., Petascale
  direct numerical simulation of blood flow on 200k cores and heterogeneous
  architectures, in: Proceedings of the 2010 ACM/IEEE International Conference
  for High Performance Computing, Networking, Storage and Analysis, IEEE
  Computer Society, 2010, pp. 1--11.

\bibitem{Quaife2014}
B.~Quaife, G.~Biros, {High-volume fraction simulations of two-dimensional
  vesicle suspensions}, Journal of Computational Physics 274 (2014) 245--267.
\newblock \href {https://doi.org/10.1016/j.jcp.2014.06.013}
  {\path{doi:10.1016/j.jcp.2014.06.013}}.

\bibitem{Kabacaoglu2018}
G.~Kabacaoǧlu, G.~Biros, {Optimal design of deterministic lateral displacement
  device for viscosity-contrast-based cell sorting}, Physical Review Fluids
  3~(12) (2018).
\newblock \href {http://arxiv.org/abs/arXiv:1805.08849v2}
  {\path{arXiv:arXiv:1805.08849v2}}, \href
  {https://doi.org/10.1103/PhysRevFluids.3.124201}
  {\path{doi:10.1103/PhysRevFluids.3.124201}}.

\bibitem{Espanol1995}
P.~Espa{\~{n}}ol,
  \href{http://link.aps.org/doi/10.1103/PhysRevE.52.1734}{{Hydrodynamics from
  dissipative particle dynamics}}, Phys. Rev. E 52~(2) (1995) 1734--1742.
\newblock \href {https://doi.org/10.1103/PhysRevE.52.1734}
  {\path{doi:10.1103/PhysRevE.52.1734}}.
\newline\urlprefix\url{http://link.aps.org/doi/10.1103/PhysRevE.52.1734}

\bibitem{Boek1997}
E.~S. Boek, P.~V. Coveney, H.~N.~W. Lekkerkerker, P.~van~der Schoot,
  {Simulating the rheology of dense colloidal suspensions using dissipative
  particle dynamics}, Physical Review E 55~(3) (1997) 3124--3133.
\newblock \href {https://doi.org/10.1103/PhysRevE.55.3124}
  {\path{doi:10.1103/PhysRevE.55.3124}}.

\bibitem{Moeendarbary2010}
E.~Moeendarbary, T.~Y. Ng, M.~Zangeneh,
  \href{http://www.worldscientific.com/doi/abs/10.1142/S1758825110000469}{{Dissipative
  Particle Dynamics in Soft Matter and Polymeric Applications — a Review}},
  International Journal of Applied Mechanics 02~(01) (2010) 161--190.
\newblock \href {https://doi.org/10.1142/S1758825110000469}
  {\path{doi:10.1142/S1758825110000469}}.
\newline\urlprefix\url{http://www.worldscientific.com/doi/abs/10.1142/S1758825110000469}

\bibitem{Quinn2011}
D.~J. Quinn, I.~V. Pivkin, S.~Y. Wong, K.-h.~H. Chiam, M.~Dao, G.~E.
  Karniadakis, S.~Suresh, {Combined simulation and experimental study of large
  deformation of red blood cells in microfluidic systems}, Annals of Biomedical
  Engineering 39~(3) (2011) 1041--1050.
\newblock \href {https://doi.org/10.1007/s10439-010-0232-y}
  {\path{doi:10.1007/s10439-010-0232-y}}.

\bibitem{Zhang2014}
P.~Zhang, C.~Gao, N.~Zhang, M.~J. Slepian, Y.~Deng, D.~Bluestein, {Multiscale
  Particle-Based Modeling of Flowing Platelets in Blood Plasma Using
  Dissipative Particle Dynamics and Coarse Grained Molecular Dynamics},
  Cellular and Molecular Bioengineering 7~(4) (2014) 552--574.
\newblock \href {https://doi.org/10.1007/s12195-014-0356-5}
  {\path{doi:10.1007/s12195-014-0356-5}}.

\bibitem{Fedosov2014}
D.~a. Fedosov, H.~Noguchi, G.~Gompper, {Multiscale modeling of blood flow: from
  single cells to blood rheology}, Biomechanics and Modeling in Mechanobiology
  13~(2) (2014) 239--258.
\newblock \href {https://doi.org/10.1007/s10237-013-0497-9}
  {\path{doi:10.1007/s10237-013-0497-9}}.

\bibitem{Lanotte2016}
L.~Lanotte, J.~Mauer, S.~Mendez, D.~A. Fedosov, J.-M. Fromental, V.~Claveria,
  F.~Nicoud, G.~Gompper, M.~Abkarian, {Red cells' dynamic morphologies govern
  blood shear thinning under microcirculatory flow conditions}, Proceedings of
  the National Academy of Sciences 113~(47) (2016) 13289--13294.
\newblock \href {https://doi.org/10.1073/pnas.1618852114}
  {\path{doi:10.1073/pnas.1618852114}}.

\bibitem{Plimpton1995}
S.~Plimpton,
  \href{http://www.sciencedirect.com/science/article/pii/S002199918571039X}{{Fast
  parallel algorithms for short-range molecular dynamics.pdf}}, Journal of
  Computational Physics 117~(1) (1995) 1--19.
\newblock \href {https://doi.org/10.1006/jcph.1995.1039}
  {\path{doi:10.1006/jcph.1995.1039}}.
\newline\urlprefix\url{http://www.sciencedirect.com/science/article/pii/S002199918571039X}

\bibitem{Tang2014}
Y.~H. Tang, G.~E. Karniadakis,
  \href{http://dx.doi.org/10.1016/j.cpc.2014.06.015}{{Accelerating dissipative
  particle dynamics simulations on GPUs: Algorithms, numerics and
  applications}}, Computer Physics Communications 185~(11) (2014) 2809--2822.
\newblock \href {http://arxiv.org/abs/1311.0402} {\path{arXiv:1311.0402}},
  \href {https://doi.org/10.1016/j.cpc.2014.06.015}
  {\path{doi:10.1016/j.cpc.2014.06.015}}.
\newline\urlprefix\url{http://dx.doi.org/10.1016/j.cpc.2014.06.015}

\bibitem{Blumers2017}
A.~L. Blumers, Y.~H. Tang, Z.~Li, X.~Li, G.~E. Karniadakis,
  \href{http://dx.doi.org/10.1016/j.cpc.2017.03.016}{{GPU-accelerated red blood
  cells simulations with transport dissipative particle dynamics}}, Computer
  Physics Communications 217 (2017) 171--179.
\newblock \href {http://arxiv.org/abs/1611.06163} {\path{arXiv:1611.06163}},
  \href {https://doi.org/10.1016/j.cpc.2017.03.016}
  {\path{doi:10.1016/j.cpc.2017.03.016}}.
\newline\urlprefix\url{http://dx.doi.org/10.1016/j.cpc.2017.03.016}

\bibitem{seaton2013dl_meso}
M.~A. Seaton, R.~L. Anderson, S.~Metz, W.~Smith, Dl\_meso: highly scalable
  mesoscale simulations, Molecular Simulation 39~(10) (2013) 796--821.

\bibitem{Schulte2015}
M.~J. Schulte, M.~Ignatowski, G.~H. Loh, B.~M. Beckmann, W.~C. Brantley,
  S.~Gurumurthi, N.~Jayasena, I.~Paul, S.~K. Reinhardt, G.~Rodgers, {Achieving
  Exascale Capabilities through Heterogeneous Computing}, IEEE Micro 35~(4)
  (2015) 26--36.
\newblock \href {https://doi.org/10.1109/MM.2015.71}
  {\path{doi:10.1109/MM.2015.71}}.

\bibitem{Rossinelli2015}
D.~Rossinelli, G.~Karniadakis, M.~Fatica, I.~Pivkin, P.~Koumoutsakos, Y.-h.
  Tang, K.~Lykov, D.~Alexeev, M.~Bernaschi, P.~Hadjidoukas, M.~Bisson,
  W.~Joubert, C.~Conti,
  \href{https://doi.org/10.1145{\%}2F2807591.2807677}{{The In-Silico
  Lab-on-a-Chip: Petascale and High-Throughput Simulations of Microfluidics at
  Cell Resolution}}, in: Proceedings of the International Conference for High
  Performance Computing, Networking, Storage and Analysis on - SC '15, ACM
  Press, 2015, pp. 1--12.
\newblock \href {https://doi.org/10.1145/2807591.2807677}
  {\path{doi:10.1145/2807591.2807677}}.
\newline\urlprefix\url{https://doi.org/10.1145{\%}2F2807591.2807677}

\bibitem{Groot1997}
R.~D. Groot, P.~B. Warren,
  \href{http://scitation.aip.org/content/aip/journal/jcp/107/11/10.1063/1.474784}{{Dissipative
  particle dynamics: Bridging the gap between atomistic and mesoscopic
  simulation}}, J. Chem. Phys. 107~(11) (1997) 4423.
\newblock \href {https://doi.org/10.1063/1.474784}
  {\path{doi:10.1063/1.474784}}.
\newline\urlprefix\url{http://scitation.aip.org/content/aip/journal/jcp/107/11/10.1063/1.474784}

\bibitem{Hoogerbrugge1992}
P.~J. Hoogerbrugge, K.~J. M.~V. A., J.~M. V.~A. Koelman, {Simulating
  Microscopic Hydrodynamic Phenomena with Dissipative Particle Dynamics},
  Europhysics Letters 19~(June) (1992) 155--160.
\newblock \href {https://doi.org/10.1209/0295-5075/19/3/001}
  {\path{doi:10.1209/0295-5075/19/3/001}}.

\bibitem{Fan2006}
X.~Fan, N.~Phan-Thien, S.~Chen, X.~Wu, T.~Y. Ng, {Simulating flow of DNA
  suspension using dissipative particle dynamics}, Physics of Fluids 18~(6)
  (2006).
\newblock \href {https://doi.org/10.1063/1.2206595}
  {\path{doi:10.1063/1.2206595}}.

\bibitem{Fedosov2010b}
D.~A. Fedosov, B.~Caswell, G.~E. Karniadakis,
  \href{http://dx.doi.org/10.1016/j.bpj.2010.02.002}{{A multiscale red blood
  cell model with accurate mechanics, rheology, and dynamics.}}, Biophysical
  Journal 98~(10) (2010) 2215--2225.
\newblock \href {https://doi.org/10.1016/j.bpj.2010.02.002}
  {\path{doi:10.1016/j.bpj.2010.02.002}}.
\newline\urlprefix\url{http://dx.doi.org/10.1016/j.bpj.2010.02.002}

\bibitem{kantor1987phase}
Y.~Kantor, D.~R. Nelson, Phase transitions in flexible polymeric surfaces,
  Physical Review A 36~(8) (1987) 4020.

\bibitem{julicher1996morphology}
F.~J{\"u}licher, The morphology of vesicles of higher topological genus:
  conformal degeneracy and conformal modes, Journal de Physique II 6~(12)
  (1996) 1797--1824.

\bibitem{Revenga1998}
M.~Revenga, I.~Z{\'{u}}{\~{n}}iga, P.~Espa{\~{n}}ol, I.~Pagonabarraga,
  {Boundary Models in DPD}, International Journal of Modern Physics C 09~(8)
  (1998) 1319--1328.
\newblock \href {https://doi.org/10.1142/S0129183198001199}
  {\path{doi:10.1142/S0129183198001199}}.

\bibitem{fedosov2008BC}
D.~A. Fedosov, I.~V. Pivkin, G.~E. Karniadakis, {Velocity limit in DPD
  simulations of wall-bounded flows}, Journal of Computational Physics 227~(4)
  (2008) 2540--2559.

\bibitem{Kotsalis2009}
E.~M. Kotsalis, J.~H. Walther, E.~Kaxiras, P.~Koumoutsakos, {Control algorithm
  for multiscale flow simulations of water} (2009).

\bibitem{Werder2005}
T.~Werder, J.~H. Walther, P.~Koumoutsakos, {Hybrid atomistic-continuum method
  for the simulation of dense fluid flows}, Journal of Computational Physics
  205~(1) (2005) 373--390.
\newblock \href {https://doi.org/10.1016/j.jcp.2004.11.019}
  {\path{doi:10.1016/j.jcp.2004.11.019}}.

\bibitem{Phillips2005}
J.~C. Phillips, R.~Braun, W.~Wang, J.~Gumbart, E.~Tajkhorshid, E.~Villa,
  C.~Chipot, R.~D. Skeel, L.~Kal{\'{e}}, K.~Schulten, {Scalable molecular
  dynamics with NAMD}, Journal of Computational Chemistry 26~(16) (2005)
  1781--1802.
\newblock \href {http://arxiv.org/abs/NIHMS150003} {\path{arXiv:NIHMS150003}},
  \href {https://doi.org/10.1002/jcc.20289} {\path{doi:10.1002/jcc.20289}}.

\bibitem{Berendsen1995}
H.~J. Berendsen, D.~van~der Spoel, R.~van Drunen, {GROMACS: A message-passing
  parallel molecular dynamics implementation}, Computer Physics Communications
  91~(1-3) (1995) 43--56.
\newblock \href {http://arxiv.org/abs/arXiv:0803.4060v1}
  {\path{arXiv:arXiv:0803.4060v1}}, \href
  {https://doi.org/10.1016/0010-4655(95)00042-E}
  {\path{doi:10.1016/0010-4655(95)00042-E}}.

\bibitem{Anderson2008}
J.~A. Anderson, C.~D. Lorenz, A.~Travesset, {General purpose molecular dynamics
  simulations fully implemented on graphics processing units}, Journal of
  Computational Physics 227~(10) (2008) 5342--5359.
\newblock \href {https://doi.org/10.1016/j.jcp.2008.01.047}
  {\path{doi:10.1016/j.jcp.2008.01.047}}.

\bibitem{pacheco2011introduction}
P.~Pacheco, An introduction to parallel programming, Elsevier, 2011.

\bibitem{Kahn1962}
A.~Kahn, {Topological Sorting of Large Networks}, Communications of the ACM
  5~(11) (1962) 558--562.
\newblock \href {https://doi.org/10.1145/368996.369025}
  {\path{doi:10.1145/368996.369025}}.

\bibitem{Tosenberger2011}
A.~Tosenberger, V.~Salnikov, N.~Bessonov, E.~Babushkina, V.~Volpert,
  \href{http://www.mmnp-journal.org/10.1051/mmnp/20116512}{{Particle Dynamics
  Methods of Blood Flow Simulations}}, Mathematical Modelling of Natural
  Phenomena 6~(5) (2011) 320--332.
\newblock \href {https://doi.org/10.1051/mmnp/20116512}
  {\path{doi:10.1051/mmnp/20116512}}.
\newline\urlprefix\url{http://www.mmnp-journal.org/10.1051/mmnp/20116512}

\bibitem{pybind11}
W.~Jakob, J.~Rhinelander, D.~Moldovan, {pybind11 — Seamless operability
  between C++11 and Python} (2016).

\bibitem{Li2008}
Z.~Li, G.~Drazer, {Hydrodynamic interactions in dissipative particle dynamics},
  Physics of Fluids 20~(10) (2008).
\newblock \href {https://doi.org/10.1063/1.2980039}
  {\path{doi:10.1063/1.2980039}}.

\bibitem{Fedosov2011}
D.~A. Fedosov, W.~Pan, B.~Caswell, G.~Gompper, G.~E. Karniadakis,
  \href{http://www.pnas.org/cgi/doi/10.1073/pnas.1101210108}{{Predicting human
  blood viscosity in silico}}, Pnas 108~(29) (2011) 11772--11777.
\newblock \href
  {https://doi.org/10.1073/pnas.1101210108/-/DCSupplemental.www.pnas.org/cgi/doi/10.1073/pnas.1101210108}
  {\path{doi:10.1073/pnas.1101210108/-/DCSupplemental.www.pnas.org/cgi/doi/10.1073/pnas.1101210108}}.
\newline\urlprefix\url{http://www.pnas.org/cgi/doi/10.1073/pnas.1101210108}

\bibitem{Backer2005}
J.~A. Backer, C.~P. Lowe, H.~C.~J. Hoefsloot, P.~D. Iedema,
  \href{http://scitation.aip.org/content/aip/journal/jcp/122/15/10.1063/1.1883163}{{Poiseuille
  flow to measure the viscosity of particle model fluids}}, Journal of Chemical
  Physics 122~(15) (2005) 154503.
\newblock \href {https://doi.org/10.1063/1.1883163}
  {\path{doi:10.1063/1.1883163}}.
\newline\urlprefix\url{http://scitation.aip.org/content/aip/journal/jcp/122/15/10.1063/1.1883163}

\bibitem{Jeffery1922}
G.~B. Jeffery, {The Motion of Ellipsoidal Particles Immersed in a Viscous
  Fluid}, Proceedings of the Royal Society of London Series A 102 (1922)
  161--179.
\newblock \href {https://doi.org/10.1098/rspa.1922.0078}
  {\path{doi:10.1098/rspa.1922.0078}}.

\bibitem{ZENG2005}
L.~Zeng, S.~Balachandar, P.~Fischer,
  \href{http://www.journals.cambridge.org/abstract{\_}S0022112005004738}{{Wall-induced
  forces on a rigid sphere at finite Reynolds number}}, Journal of Fluid
  Mechanics 536 (2005) 1--25.
\newblock \href {http://arxiv.org/abs/arXiv:1011.1669v3}
  {\path{arXiv:arXiv:1011.1669v3}}, \href
  {https://doi.org/10.1017/S0022112005004738}
  {\path{doi:10.1017/S0022112005004738}}.
\newline\urlprefix\url{http://www.journals.cambridge.org/abstract{\_}S0022112005004738}

\bibitem{Fedosov2010}
D.~A. Fedosov, {Multiscale Modeling of Blood Flow and Soft Matter}, Ph.D.
  thesis, Brown University (2010).
\newblock \href {https://doi.org/10.1080/01635580802395717}
  {\path{doi:10.1080/01635580802395717}}.

\bibitem{Karabacak2014a}
N.~M. Karabacak, P.~S. Spuhler, F.~Fachin, E.~J. Lim, V.~Pai, E.~Ozkumur, J.~M.
  Martel, N.~Kojic, K.~Smith, P.-i. Chen, J.~Yang, H.~Hwang, B.~Morgan,
  J.~Trautwein, T.~A. Barber, S.~L. Stott, S.~Maheswaran, R.~Kapur, D.~a.
  Haber, M.~Toner,
  \href{http://www.ncbi.nlm.nih.gov/pubmed/24577360}{{Microfluidic, marker-free
  isolation of circulating tumor cells from blood samples.}}, Nature protocols
  9~(3) (2014) 694--710.
\newblock \href {https://doi.org/10.1038/nprot.2014.044}
  {\path{doi:10.1038/nprot.2014.044}}.
\newline\urlprefix\url{http://www.ncbi.nlm.nih.gov/pubmed/24577360}

\bibitem{Huang2004a}
L.~R. Huang, E.~C. Cox, R.~H. Austin, J.~C. Sturm, {Continuous particle
  separation through deterministic lateral displacement.}, Science (New York,
  N.Y.) 304~(5673) (2004) 987--990.
\newblock \href {https://doi.org/10.1126/science.1094567}
  {\path{doi:10.1126/science.1094567}}.

\end{thebibliography}




%%   \bibitem{key}figures.
%%


% \end{thebibliography}

\end{document}